\documentclass{emulateapj}
\usepackage{mathrsfs,amsmath}
\begin{document}
%------------------------------------------------------------------------------

\title{Conservative GRMHD Simulations of Moderately Thin, Tilted Accretion Disks}
\shorttitle{GRMHD Simulation of Thin, Tilted Disk}
\shortauthors{Teixeira et al.}
\author{Danilo Morales Teixeira}
\affil{Instituto de Astronomia, Geof\'isica e Ci\^encias Atmosf\'ericas, Universidade de S\~ao Paulo, S\~ao Paulo, SP 05508-090, Brazil}
\email{danilo@astro.iag.usp.br}
\author{P. Chris Fragile\footnote{KITP Visiting Scholar, Kavli Institute for Theoretical Physics, Santa Barbara, CA.}}
\affil{Department of Physics \& Astronomy, College of Charleston, Charleston, SC 29424, USA}
\author{Viacheslav V. Zhuravlev}
\affil{Sternberg Astronomical Institute, M.V. Lomonosov Moscow State University, Universitetskij pr. 13, 119992 Moscow, Russia}
\and
\author{Pavel B. Ivanov}
\affil{Astro Space Centre, P. N. Lebedev Physical Institute, 84/32 Profsoyuznaya Street, 117810 Moscow, Russia}

\begin{abstract}
This paper presents our latest numerical simulations of accretion disks that are misaligned with respect to the rotation axis of a Kerr black hole. In this work we use a new, fully conservative version of the Cosmos++ general relativistic magnetohydrodynamics (GRMHD) code, coupled with an ad hoc cooling function designed to control the thickness of the disk. Together these allow us to simulate the thinnest tilted accretion disks ever using a GRMHD code. In this way, we are able to probe the regime where the dimensionless stress and scale height of the disk become comparable. We present results for both prograde and retrograde cases. The simulated prograde tilted disk shows no sign of Bardeen-Petterson alignment even in the innermost parts of the disk. The simulated retrograde tilted disk, however, does show modest alignment. The implication of these results is that the parameter space associated with Bardeen-Petterson alignment for prograde disks may be rather small, only including very thin disks. Unlike our previous work, we find no evidence for standing shocks in our simulated tilted disks. We ascribe this to the combination of small black hole spin, small tilt angle, and small disk scale height in these simulations. We also add to the growing body of literature pointing out that the turbulence driven by the magnetorotational instability in global simulations of accretion disks is not isotropic. Finally, we provide a comparison between our moderately thin, untilted reference simulation and other numerical simulations of thin disks in the literature. 
\end{abstract}
\keywords{accretion, accretion disks --- black hole physics --- relativistic processes}
\maketitle

\section{Introduction}
\label{sec:intro}

Twisted and tilted accretion disks have been investigated for nearly 40 years, starting from the seminal paper of \citet{Bardeen75}. Until relatively recently such disks were studied in the framework of different theoretical schemes, where the disk's tilt and twist have been described by two Euler angles, $\beta (r, t)$ and $\gamma (r,t)$, characterizing the inclination and precession of each disk's ring.
Dynamical equations for $\beta (r, t)$ and $\gamma (r,t)$ have then been derived under various assumptions.
 
Stationary configurations have also been derived.  In their original work, \citet{Bardeen75} proposed such a configuration in which the inclination angle decreased toward the black hole ($d\beta /dr > 0$), and the disk, accordingly, tended to align with the black hole equatorial plane -- the so-called Bardeen-Petterson effect. It was shown, however, by \citet{Papaloizou83} that the approach of \citet{Bardeen75} was quantitatively incorrect.
\citet{Papaloizou83} and later \citet{Kumar85} found new stationary solutions, which showed that the Bardeen-Petterson effect holds, but the characteristic radial scale of the disk's alignment decreases when the Shakura-Sunyaev parameter $\alpha$ gets smaller, contrary to the original \citet{Bardeen75} claim. Later, a time dependent relaxation to a stationary configuration was investigated by \citet{Kumar90}. This was later generalized by \citet{Ogilvie99} to take into account non-linear effects in the disk's inclination angle $\beta (r, t)$. However, when the $\alpha$ parameter becomes smaller than the ratio of the disk's semi-thickness to its radius, the \citet{Papaloizou83} theory needs to be modified to take into account both sonic effects \citep{Papaloizou95} and the effect of Einstein apsidal precession \citep{Ivanov97}. In this regime, the stationary configurations do not necessarily show alignment of the disk with the black hole equatorial plane, as was shown by \citet{Ivanov97} and confirmed by \citet{Lubow02} and \citet{Zhuravlev11}.

The validity of the Bardeen-Petterson effect in numerical studies is also ambiguous. In the smoothed-particle hydrodynamic (SPH) simulations done by \citet{Nelson00}, the effect was observed. Those simulations, however, had a large effective viscosity and could not, therefore, test the case of small viscosity, where the analytic theory predicts that the Bardeen-Petterson effect may be invalid. In a similar way, the SPH simulations of \citet{Lodato07,Lodato10} confirmed the analytic predictions of \citet{Papaloizou83}\footnote{Note that \citet{Lodato10} also found some deviation from the simple linear theory of tilted disks in situations where the gradient of the disk's tilt is sufficiently large. Deviations of this type are not expected in this work, however, since we deal only with rather small gradients.}, but were subject to the same assumptions, namely that the physics is adequately described via the normal Newtonian equations plus a torque term, that the ``viscosity'' is isotropic, and that the \citet{Shakura73} parameter, $\alpha$, is constant.  Each of these assumptions is now known to be incorrect; hence the need for more realistic numerical simulations.

More recently, \citet{Sorathia13} performed an MHD simulation of a moderately thin tilted disk and found that the  Bardeen-Petterson alignment also holds in their case. However, these authors also modeled the black hole gravitational field via two classical forces: one describing a point-like Newtonian potential and the other describing a gravitomagnetic torque. However, as was discussed by \citet{Ivanov97}, in order to obtain a tilted disk configuration where the  
Bardeen-Petterson effect is violated, one must take into account relativistic corrections to the point-like Newtonian potential, in particular the Einstein precession of the line of apsides. 

On the other hand, in the fully relativistic numerical simulations by \citet{Fragile07, Fragile09a, Fragile09b}, the Bardeen-Petterson effect has not been observed. However, these authors considered only the case of geometrically thick accretion disks, which are not as easily compared with analytic theory.

In this paper we present the first fully relativistic simulations of tilted thin accretion disks ($H/r=0.08$), where the effective viscosity is induced by the development of the magnetorotational instability, and examine whether the Bardeen-Petterson effect holds.  We also include reference simulations of untilted accretion disks; as these are still among only a small selection of such published simulations \citep[others include][]{Shafee06, Reynolds08, Noble09, Noble10, Penna10}, they are worthy of mention, in and of themselves.

In our companion paper \citep[][hereafter referred to as Paper 2]{Zhuravlev14}, we compare the numerical results discussed in this paper with a fully relativistic semi-analytic model of a tilted disk, based on the time-dependent formalism introduced in \citet{Zhuravlev11}.  The semi-analytic model allows us to clarify the physical meaning of the simulation results.  Both the simulations and models show the same qualitative behavior in the disk. Namely, when the disk rotates in the same sense as the black hole (i.e. prograde), Bardeen-Petterson alignment is absent and the disk's inclination angle actually grows slightly toward the black hole. This can be understood in terms of a standing bending wave. 
In the case when the disk and  black hole rotations are opposite (i.e. retrograde), we observe instead a modest alignment of the disk toward the black hole equatorial plane, which may be interpreted as the  Bardeen-Peterson effect. Note that this is the first fully relativistic simulation in which this type of alignment is observed.

In addition to our focus on the effects of tilt, we also consider whether the often-made assumption of an isotropic viscosity is valid. We find that although our simulations yield a reasonable value for $\alpha$, defined in terms of the ``horizontal'' $r-\phi$ components of the stress and rate-of-strain tensors, the same is not true for the other components. The anisotropy of the turbulent stresses may be important for many theoretical models of tilted disks, as they have often relied on the assumption of isotropy. 
 
The presentation of our work is organized as follows: In Section \ref{sec:method} we describe the numerical simulations, including an important discussion about resolution.  In Section \ref{sec:untilted}, we make comparisons between our {\em untilted} simulations and earlier, similar simulations by other authors.  Since simulations of moderately thin disks are still relatively new and scarce in the literature, these additional comparisons seem worthwhile.  In Section \ref{sec:tilted}, we get to the main result of this paper, which is a presentation of the results of our {\em tilted} simulations.  
Closely connected to our main results is the discussion in Section \ref{sec:viscosity} of the anisotropic effective stress tensor that we get from the magnetorotational instability in our simulations.  Finally, we end in Section \ref{sec:conclusion} with some discussion and conclusions. 

We use, hereafter, the natural system of units, setting the speed of light and gravatational constant to unity.  We adopt the $(-,+,+,+)$ metric signature.

\section{Numerical Simulations}
\label{sec:method}

In this work, we present seven numerical simulations, covering different resolutions, tilted and untilted cases, prograde and retrograde disks, and testing different procedures for introducing the tilt.  The main difference between these simulations and our earlier work \citep{Fragile07,Fragile09a} is that here we use the new, fully-conservative, high-resolution shock capturing (HRSC) method of the {\em Cosmos++} astrophysical fluid dynamics code \citep{Fragile12}, plus an {\it ad-hoc} cooling function to control the scale height of the disk \citep{Fragile12b}.  In all of the simulations presented here, we set the target relative scale height to $\delta = H/r =  0.08$, making these the thinnest tilted disks ever simulated using GRMHD.  Our motivation for choosing this value, as well as $\vert a_* \vert = 0.1$ for the black hole spin, are twofold: first, these values should be reasonably small to facilitate comparison with our semi-analytic model, as discussed in Paper 2; second, we want values that give us a reasonable chance to capture the spatial scales associated with Bardeen-Petterson disks within our computational domain. This also motivates how long we run our simulations, since we wish, at a minimum, to cover the relevant timescales in the inner parts of the disk.  

Following our previous GRMHD simulations, we use Kerr-Schild coordinates $(t, r, \theta, \phi)$ as the principal coordinate system in our numerical work. To introduce the tilt, we make a rotation of the coordinate system by an amount $\beta_0$ about the y-axis, as in equation (11) of \citet{Fragile07}. This results in a change of the spherical angles $(\theta, \phi)\rightarrow (\vartheta, \varphi)$ such that initial disk mid-plane always coincides with the equatorial plane of the tilted coordinates, $\vartheta =\pi/2$.

\subsection{Initialization}
\label{sec:setup}

There are many possible starting configurations that one could consider. Motivated by our own earlier numerical work, we have chosen to initialize the simulations with an axisymmetric torus orbiting around a rotating black hole \citep{Abramowicz78}, following the procedure described in \citet{Chakrabarti85}. Through the action of turbulence generated by the magnetorotational instability \citep[MRI][]{Balbus91}, this torus rapidly evolves into an accretion disk.  The hope is that the resulting disk will settle to the Bardeen-Petterson solution. Another option would be to try to start the simulation from one of the stationary, tilted-disk solutions in the literature \citep[e.g.][]{Papaloizou83,Kumar85} and see if the simulation retains this solution. We plan to consider this in future work.

The initial torus configuration is specified by its inner radius ($r_\mathrm{in}$), the radius of its pressure maximum ($r_\mathrm{cen}$), the black hole spin ($a_* = a/M$), and a parameter ($q$) that is used to define the angular momentum distribution 
\begin{equation}
l=-\frac{u_\varphi}{u_t}=k\Omega^{1-2/q} ~,
\end{equation}
where
\begin{equation}
\Omega =  u^\phi/u^t=-\frac{g_{t\varphi} + lg_{tt}}{g_{\varphi\varphi}+lg_{t\varphi}} ~.
\end{equation}
Here $u_\mu$ is the covariant 4-velocity of the gas, $g_{\mu\nu}$ are the Kerr-Schild metric coefficients in terms of the tilted coordinates, and $k$ is a coefficient that is fixed by the requirement that $l = l_\mathrm{Kep}$ at $r=r_\mathrm{cen}$.  Notice that we are not accounting for the effects of tilt in initializing our torus, rather assuming that the equatorial coordinate plane is equivalent to the black hole equatorial plane, which of course it is not whenever $\beta_0 \ne 0$.  However, since the torus just serves as a convenient starting point and will be unstable because of the MRI anyway, this discrepancy is not of significant concern.  Furthermore, as we explain below, most of the simulations start from an untilted ($\beta_0 = 0$) configuration anyway.

In this work we fix $r_\mathrm{in} = 12 M$, $\vert a_* \vert= 0.1$, and $q=1.6$.  We mention here that we will often use the orbital period of a test particle at $r_\mathrm{cen}$, i.e. $t_\mathrm{orb} = 2\pi/\Omega(r_\mathrm{cen})$, as a convenient time unit when discussing our results.
The specific internal energy of the gas is fixed by the spacetime and knowledge of $l(r=r_\mathrm{in})$ \citep[see e.g.][and references therein for details]{Fragile07}.  This fixes the gas density, $\rho \propto \epsilon^{1/(\Gamma-1)}$, up to an arbitrary constant, where we assume a $\Gamma = 5/3$ polytrope.  Finally, the gas pressure is fixed by assuming an ideal gas equation of state (EOS),
\begin{equation}
P_g = (\Gamma-1)\rho\epsilon ~.
\end{equation}

In order to seed the MRI we add a weak initial magnetic field, composed of a single set of poloidal loops that follow the isodensity contours of the torus, such that initially $\beta_\mathrm{mag}=P_g/P_B \ge10$ everywhere in the torus, where $P_g$ and $P_B$ are the gas and magnetic pressures, respectively.  This field configuration does not contain enough magnetic flux to lead to a magnetically-arrested state over the course of the simulation \citep{McKinney12a}; tilted accretion disks in such a state are considered in \citet{McKinney12b}.

In the background region where the torus solution does not apply, we set up a rarefied non-magnetic plasma accreting into the black hole \citep{Komissarov06}. The density and internal energy density of this gas are given by $\rho=10^{-4}\rho_{\mathrm{max},0} (r/M)^{-2.7}$ and $e = \rho\epsilon = 10^{-6} \rho_{\mathrm{max},0} (r/M)^{-2.7}$, where $\rho_{\mathrm{max},0}$ is the maximum initial density in the torus.  Because there is a free scale in the problem, the absolute numerical value of this density is irrelevant.  These profiles also serve as numerical floors on the values of $\rho$ and $e$ during the evolution of the simulation.  The initial radial velocity profile of this background gas is given by
\begin{equation}
V^r=\frac{g^{tr}}{g^{tt}}\left[1-\left(\frac{M}{r}\right)^4\right] ~.
\end{equation}

The simulations are carried out on a uniformly spaced grid of spatial coordinates \{$x_1$, $x_2$, $\varphi$\}.  All curvature, both real and coordinate, is handled via the metric.  This procedure relies on the following transformations between the grid coordinates $x_1$ and $x_2$ and the corresponding Kerr-Schild spatial ones:
\begin{equation}
r(x_1) = r_H e^{x_1} 
\end{equation}
and 
\citep{Noble10}
\begin{equation}
\vartheta(x_2)=\frac{\pi}{2}\left[1+(1-\varepsilon)(2x_2-1)+\varepsilon(2x_2-1)^n\right] ~,
\end{equation}
where $n$ is any positive, odd integer (taken to be 9 in our case) and $\varepsilon = 0.3$ is the amplitude of the nonlinear term.  Note that we do not use a cut out region around the pole.  Because we use horizon-penetrating Kerr-Schild coordinates, we are able to place the inner radial boundary of the grid inside the black hole event horizon, $r_H = M (1+\sqrt{1-a^2_*})$ ($r_H= 1.99 M$ for $a_* = 0.1$), thus isolating it from the physical domain of interest.  The inner radial boundary is at $0.9 r_H$, giving 7 zones inside the event horizon for our high-resolution simulations (and 3 for the medium ones).  The outer boundary is placed at $100 M$.  We use outflow boundary conditions at both the inner and outer radial boundaries, reflecting boundaries on the pole, and periodic boundaries in the azimuthal direction.

\subsection{Resolution}
\label{sec:resolution}

Before continuing, we need to address the significant drawback of us trying to simulate the thinnest tilted disks ever using GRMHD -- numerical resolution becomes a major issue.  Thin accretion disks are challenging in general.  Resolving the vertical scale height of the disk while simultaneously maintaining comparable resolution in all three spatial dimensions (a requirement of most grid-based numerical techniques), requires many more zones in the radial and azimuthal directions, roughly $N_r \sim N_\phi \sim N_z/\delta$.  For a thin disk, with $\delta \ll 1$, this can be prohibitive.  Further, in the case of a titled disk, there are no symmetries to the problem that can be exploited; the full three-dimensional domain must be treated.

To keep the computational expense of this project reasonable, we exploit the fact that we are considering a very small tilt and a moderately thin disk by concentrating most of the $\vartheta$ zones very near the mid-plane.  Even so, our simulations are admittedly very under-resolved with respect to the MRI.  In the vertical direction we have approximately 3 zones per MRI wavelength ($Q_\vartheta \ge 3$), and in the azimuthal direction we have about 9 ($Q_\varphi \ge 9$) for our high-resolution simulations, at relevant times ($t \ge 8 t_\mathrm{orb} \approx 8300 M$) over the radial range ($10 \le r/M \le 50$), where
\begin{equation}
Q_x=\frac{\lambda_{x,MRI}}{\Delta _x} ~,
\end{equation}
$\Delta_\vartheta=rd\vartheta$, $\Delta _\varphi=r\sin\vartheta d\varphi$, and the fastest growing MRI wavelength is given by
\begin{equation}
\lambda _{x,\mathrm{MRI}}=2\pi\frac{v_{x,A}}{\Omega},
\end{equation}
where $v_{x,A}=\sqrt{b_xb^x/(\rho + \rho\epsilon + P_g + 2 P_B)}$ is the Alfv\'en speed in the direction of interest; $b^{i} \equiv u_\beta {^*F^{i\beta}}$ are components of magnetic field defined in the co-moving frame, where ${^*F^{\alpha\beta}}$ is the dual of the Faraday tensor; $\epsilon$ is the co-moving gas thermal energy per unit of mass;  and $P_B = b^\alpha b_\alpha/2$ is the magnetic pressure.  The $Q$ values improve somewhat inside of $r = 10 M$.

Our modest $Q$ values mean that the MRI turbulence is not fully developed in these simulations \citep[see][for discussions of MRI resolution in numerical simulations]{Hawley11,McKinney12a,Hawley13}.  However, we make the following arguments for why our results are still meaningful:  First, although this likely means that our effective values for $\alpha$ may be smaller than they would be in better resolved simulations, we, nevertheless, have measurable effective viscosities acting in our simulations, and we can meaningfully discuss the response of the disks to them.  Second, it has been shown in previous studies \citep[e.g.][]{Fragile08,Sorathia13} that the main responses within tilted disks are hydrodynamic, not MHD.  Therefore, despite the poor resolution of the MRI, the leading order effects in tilted disks are still captured.  Additionally, in Section \ref{sec:untilted} we show that our untilted simulations reproduce many of the key features seen in other simulations of thin accretion disks.  Finally, in this and previous studies, we have performed our simulations at different resolutions and confirmed that we find substantially the same behavior at all resolutions.

Nevertheless, there may be one important consequence of our poor resolution of the MRI: the lack of well-developed turbulence may allow the disk to precess more nearly as a solid body than would be the case otherwise.  In comparing their hydrodynamic and MHD simulations of Lense-Thirring precession, \citet{Sorathia13} found that the MHD simulations showed less solid-body precession.  They argued that the turbulence in the MHD simulation acted to break up the coupling between neighboring rings that allows for solid-body precession.  This may also play a role in how effectively the disk can align with the black hole.

\subsection{Evolution}

Our numerical scheme is explained in more detail in \citet{Anninos05,Fragile12}.  Here, we provide only a brief summary of the most relevant details.  First, we solve the following set of coupled GRMHD equations:
\begin{equation}
\partial_t D + \partial_i (DV^i) = 0 ~,  \label{eqn:de} 
\end{equation}
\begin{equation}
 \partial_t {\cal E} + \partial_i (-\sqrt{-g}~T^i_t) =
      -\sqrt{-g}~T^\kappa_\lambda~\Gamma^\lambda_{t\kappa} + \sqrt{-g}~\Lambda u_t ~,
    \label{eqn:en}
\end{equation}
\begin{equation}
 \partial_t {\cal S}_j + \partial_i (\sqrt{-g}~T^i_j) =
      \sqrt{-g}~T^\kappa_\lambda~\Gamma^\lambda_{j\kappa} - \sqrt{-g}~\Lambda u_j ~,
    \label{eqn:mom}
\end{equation}
\begin{equation}
 \partial_t \mathcal{B}^j + \partial_i (\mathcal{B}^j V^i - \mathcal{B}^i V^j) =
     0 ~, 
      \label{eqn:ind} 
\end{equation}
where
\begin{equation}
T^{\alpha \beta} = (\rho + \rho\epsilon + P_g + 2 P_B) u^\alpha u^\beta + (P_g + P_B)g^{\alpha \beta} - b^\alpha b^\beta
\label{eq:tmunu}
\end{equation}
is the usual MHD stress-energy tensor, $D = W\rho$ is the generalized fluid density, ${\cal E} = - \sqrt{-g} T^t_t$ is the total energy density, ${\cal S}_j = \sqrt{-g} T^t_j$ is the covariant momentum density, $\mathcal{B}^j = \sqrt{-g} B^j$ is the boosted magnetic field three-vector, $B^i = {^*F^{\alpha i}}$, $W = \sqrt{-g}u^t$ is the generalized boost factor, and $g$ is the metric determinant.  While most of these variables are defined at the respective cell centers, the magnetic field components, $\mathcal{B}^j$, are staggered, residing at the respective cell faces.  This facilitates application of our constrained transport procedure for maintaining a divergence-free field, as described in \citet{Fragile12}.

The relative scale height of the disk ($\delta$) is controlled through a cooling function of the form \citep{Noble09}
\begin{equation}
\Lambda=\Omega_*\rho\epsilon[Y-1+\vert Y-1\vert]^{1/2} ~,
\end{equation}
where $Y=(\Gamma-1)\epsilon/T_*$ is the ratio of the actual to the target temperature
\begin{equation}
T_*=\frac{\pi}{2}\left[\delta_* r\Omega_* \right]^2 ~,
\end{equation}
\(\Omega_*(r)=M^{1/2}/(r^{3/2}+aM^{1/2})\) is the relativistic orbital frequency, and $\delta_*$ is the target scale height (set to 0.08 throughout this work).

The basic parameters of each simulation are given in Table \ref{tab:params}.  The naming convention is in keeping with our previous work and such that the first number indicates the dimensionless black hole spin, in units of tenths; the second number indicates the tilt of the disk, in units of degrees; if followed by the letter ``r'', then the simulation is retrograde; the next letter indicates the resolution, either ``m'' for medium or ``h'' for high; and finally, the letter ``t'' indicates a simulation that started with a tilt.  Most cases started with an untilted torus.  Simulations 10m, 10rm, and 10h are run to their final termination time, $t_\mathrm{stop}$, in this untilted configuration.  These simulations serve as reference simulations, allowing us to better isolate the effects of tilt.  They are also used as background models for the semi-analytic approach, as described in Paper 2.  For simulations 110m, 110rm, and 110h, we restart simulations 10m, 10rm, and 10h, respectively, from a time $t = 2 t_\mathrm{orb} \approx 2100 M$, but with a tilt of $\beta_0 = 10^\circ$.  We then allow these simulations to run out to a cumulative evolution time of $t_\mathrm{stop}$ ($\approx 34 t_\mathrm{orb} \approx 35000 M$ for 110m, $\approx 22 t_\mathrm{orb} \approx 26000 M$ for 110rm, and $\approx 12.5 t_\mathrm{orb} \approx 13000 M$ for 110h).  This means that the MRI is already established and the disk is accreting before the tilt is introduced.  For the final simulation, 110mt, we use the procedure from \citet{Fragile07}, whereby the tilt is introduced immediately after setting up the initial torus.  Thus, even the early evolution of this disk includes the effects of tilt.  The simulations 10rm and 110rm are our retrograde cases (the black hole spins counter to the disk angular momentum).  These are included since retrograde cases are always expected to show a tendency toward (counter-) alignment.

\begin{deluxetable}{cccccc}
\tablecaption{Simulation Parameters\tablenotemark{a} \label{tab:params}}
\tablewidth{0pt}
\tablehead{
\colhead{Simulation} & \colhead{Resolution} & $r_\mathrm{cen}/M$ & $t_\mathrm{orb}/M$ & \colhead{Tilt ($\beta_0$)} & \colhead{$t_\mathrm{stop}/t_\mathrm{orb}$}
}
\startdata
10m & $128 \times 48 \times 96$ & 30 & 1033 & $0^\circ$ & 34 \\
10rm & $128 \times 48 \times 96$ & 33.2 & 1202 & $0^\circ$ & 22 \\
110m & $128 \times 48 \times 96$ & 30 & 1033 & $10^\circ$ & 34 \\
110rm & $128 \times 48 \times 96$ & 33.2 & 1202 & $10^\circ$ & 22 \\
110mt & $128 \times 48 \times 96$ & 30 & 1033 & $10^\circ$ & 22 \\
10h & $256 \times 96 \times 192$ & 30 & 1033 & $0^\circ$ & 12.5 \\
110h & $256 \times 96 \times 192$ &30 & 1033 & $10^\circ$ & 12.5 \\
\enddata
\tablenotetext{a}{The following parameters remain fixed for all simulations: $r_\mathrm{in} = 12 M$, $\vert a_* \vert= 0.1$, and $q=1.6$.}
\label{table1}
\end{deluxetable}

\section{Comparison with Previous Untilted Simulations}
\label{sec:untilted}

Since these are the first 3D global, GRMHD simulations we have done of thin disks, before considering our tilted simulations, we provide a brief comparison of our untilted ones with earlier GRMHD simulations of comparable thickness, notably the work of \citet{Noble10} and \citet{Penna10}.  Global GRMHD simulations of thin disks are also still relatively new, so it is worthwhile to provide these extra data.

\begin{figure}
\begin{center}
\includegraphics[width=0.9 \columnwidth,angle=0]{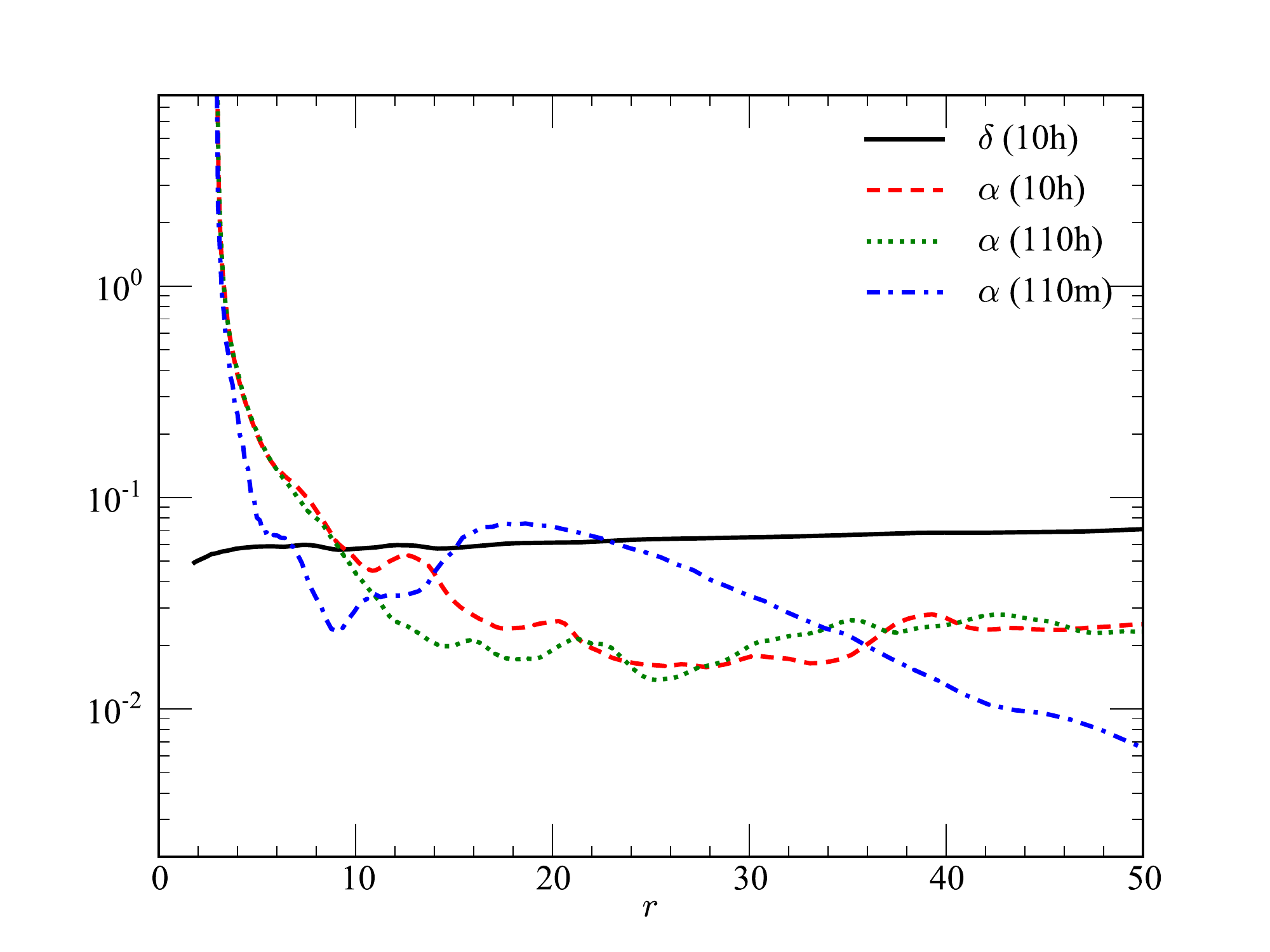}
\caption{Comparison of density-weighted shell averages of scale height, $\langle\delta\rangle$, and dimensionless stress, $\overline{\alpha_{\hat{r}\hat{\phi}}}$, for simulations 10h, 110h, and 110m, time-averaged over the final $2.5 t_\mathrm{orb}$ of each simulation.  We see that the cooling function does a reasonably good job of maintaining the desired scale height throughout the disk, which is within a factor of a few of the typical effective $\alpha$.  
}
\label{fig:alpha}
\end{center}
\end{figure}

First, to demonstrate that our cooling function is performing as intended, and that we indeed have a moderately thin disk, in Figure \ref{fig:alpha} we plot a time-averaged, radial profile of the disk scale height, $\langle\delta\rangle$.  To estimate it, we use the following expression from \citet{Penna10}
\begin{equation}
\langle\delta\rangle=\sqrt{\frac{\int d\vartheta d\varphi\sqrt{-g}\rho^2 \left(\vartheta -\frac{\pi}{2}\right)^2}{\int d\vartheta d\varphi\sqrt{-g}\rho^2}} ~,
\label{delta}
\end{equation}
where the scale height is weighted by the square of the density and the integrals are over all cells within a given radial shell.  This is a different weighting than we have used in earlier works \citep[e.g.][]{Fragile07}.  This new expression is preferred because it gives a better agreement with the target $\delta_*$ that we use in our cooling function.   Note, however, that the value we get from this estimate of $\langle\delta\rangle$ does not necessarily agree quantitatively with other estimates.  For example, one might consider $\delta = t_\mathrm{dyn}/t_\mathrm{cs}$ (see Section \ref{sec:timescales} for definitions), which comes from the relationship for the vertically-integrated sound speed, $c_s = H\Omega$.  Figure \ref{fig:timescales} would then suggest $\delta \approx 0.12$.

Figure \ref{fig:alpha} also includes the time-averaged profile of the effective horizontal viscosity, $\overline{\alpha}_{\hat{r}\hat{\phi}} = -3\langle W_{\hat{r}\hat{\phi}}\rangle\langle\Omega\rangle /(4\langle S_{\hat{r}\hat{\phi}}\rangle (\langle P_g \rangle + \langle P_B \rangle))$ (see Section \ref{sec:viscosity} for a complete description of how we define $\alpha_{\hat{i}\hat{j}}$).  Here, a normal density weighting is used: 
\begin{equation}
\langle X(r) \rangle=\frac{\int d\vartheta d\varphi\sqrt{-g}\rho X(r,\vartheta,\varphi)}{\int d\vartheta d\varphi\sqrt{-g}\rho} ~.
\end{equation}
This profile of $\overline{\alpha_{\hat{r}\hat{\phi}}}$ shows similar properties to other such profiles extracted from GRMHD simulations \citep[see, e.g.,][]{Penna13}, in that it has a strong radial dependence at small radii and then asymptotes to a value between $0.01-0.1$ at large radii.  One difference is that our profile does not turn over and begin decreasing inside the innermost stable circular orbit (ISCO), as it does in \citet{Penna13}, but we must emphasize that we are using a different definition of $\alpha_{\hat{r}\hat{\phi}}$.  Most other work, including our own previous papers, have defined $\alpha_{\hat{r}\hat{\phi}}$ only in terms of the MHD stress tensor in the co-moving frame, normalized to the total pressure, i.e. $\alpha_{\hat{r}\hat{\phi}} = W_{\hat{r}\hat{\phi}}/P_\mathrm{Tot}$.  This simplification relies on the fact that for purely Keplerian disks, $S_{\hat{r}\hat{\phi}} \propto \Omega$.  Here, if we use the same definition, we get profiles that look very similar to profiles from other GRMHD simulations.  However, it is not obvious that $S_{\hat{r}\hat{\phi}}$ should be proportional to $\Omega$ in the same way for tilted disks.  Furthermore, one of our goals in Section \ref{sec:viscosity} is to test whether $W_{\hat{i}\hat{j}} \propto S_{\hat{i}\hat{j}}$, a common assumption in most previous analytic work on tilted accretion disks.  Therefore, we use the more general expression above.

\begin{figure}
\begin{center}
\includegraphics[width=0.8 \columnwidth,angle=0]{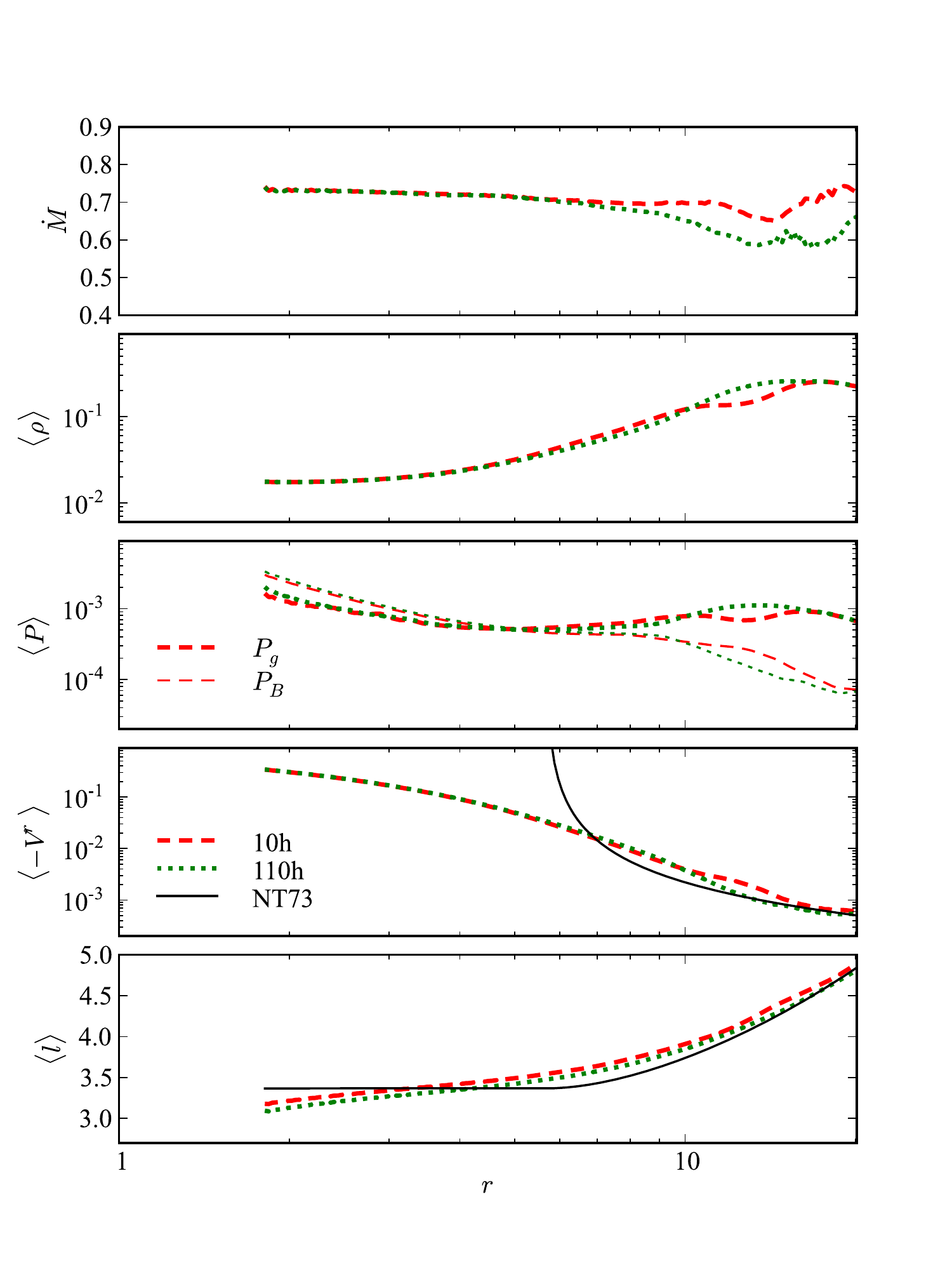}
\caption{Time-averaged gas mass accretion rate (arbitrary units) ({\em top} panel), plus density-weighted radial shell averages of the gas density (arbitrary units)({\em second} panel), gas ({\em thick} lines) and magnetic ({\em thin} lines) pressures (arbitrary units) ({\em third} panel), inflow velocity ({\em fourth} panel), and specific angular momentum ({\em bottom} panel) for the untilted (10h) and tilted (110h) high-resolution simulations.  Data are time averaged over the interval $10 \le t/t_\mathrm{orb} \le 12.5$.  The {\em thin, black} lines in the {\em third} and {\em fourth} panels show the expectations of the \citet{Novikov73} model, assuming $\alpha = 0.1$ and $\delta = 0.08$.}
\label{fig:profiles}
\end{center}
\end{figure}

In Figure \ref{fig:profiles}, we present additional time-averaged radial profiles for simulations 10h and 110h.  The {\em top} panel presents the time-averaged, radial mass accretion rate, 
\begin{equation}
\dot{M} = - \int \sqrt{-g} \rho u^r d\vartheta d\varphi~.
\end{equation}
This shows that both simulations have achieved a reasonable inflow equilibrium inside of $r \approx 10 M$, with the tilted simulation accreting at a slightly higher rate, consistent with our earlier work \citep{Fragile07,Fragile08,Generozov14}.  There is, however, a slight rise in $\dot{M}$ inside the ISCO, which we comment on below. The {\em second} and {\em third} panels show the radial profiles of the shell-averaged gas density, $\langle \rho \rangle$, and density-weighted, shell-averaged gas and magnetic pressures, $\langle P_g \rangle$ and $\langle P_B \rangle$, respectively. The {\em fourth} panel shows the density-weighted, shell-averaged radial velocity profile, $\langle -V^r \rangle$.  This profile is almost exactly the same as the comparable profile in Figure 6 of \citet{Penna10}, and follows fairly closely the predictions of \citet[][NT73]{Novikov73} outside of the ISCO.  Finally, the {\em bottom} panel shows the density-weighted, shell-averaged specific angular momentum profile, $\langle l \rangle$.  Again, this plot is very similar to the comparable plot in Figure 14 of \citet{Noble10}.  It also tracks fairly closely the model of NT73.  The main discrepancy is that $\langle l \rangle$ continues to drop inside the ISCO, indicating that stresses are still acting to extract angular momentum from the flow, in contrast to the assumption of NT73.  This same behavior was noted in \citet{Noble10} and \citet{Penna10}.  The results of \citet{Penna10} suggest that this discrepancy would become negligible if the disk thickness were to approach zero.  Their results also suggest that the discrepancy would be smaller if we had initialized our problem with multiple, alternating islands of poloidal magnetic field loops instead of a single set.  Additional radial profiles of $\alpha_{\hat{r}\hat{\phi}}$, $\delta$, the surface density $\Sigma$, and $u^r$ at different times, for both the high and medium resolution untilted simulations (10h and 10m), are provided in Paper 2.  There the profiles serve as inputs to our semi-analytic models, thus allowing the models to capture the effects of both temporal and spatial variability.  

\begin{figure}
\begin{center}
\includegraphics[width=0.9 \columnwidth,angle=0]{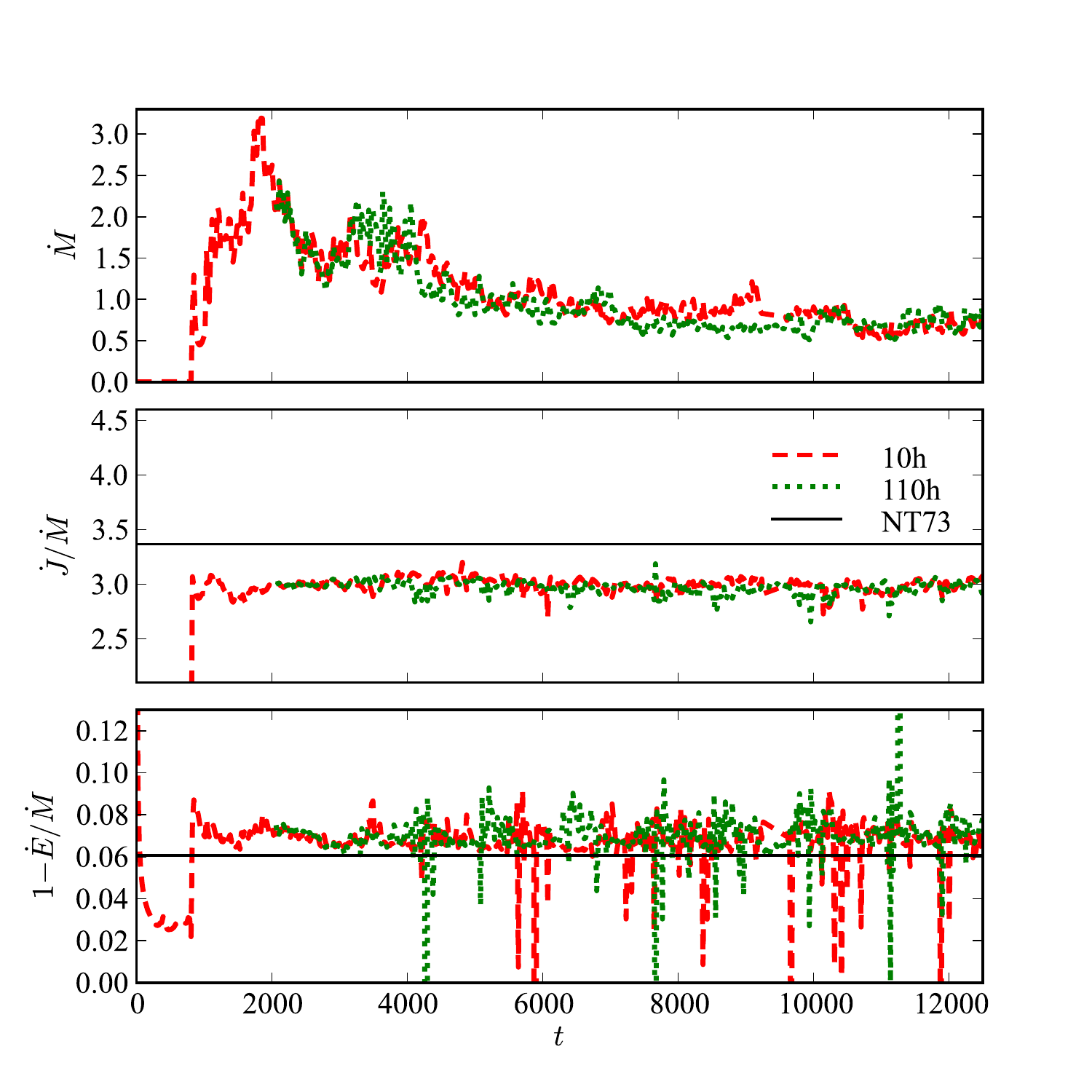}
\caption{Time dependence of the event-horizon mass and specific angular momentum fluxes and the nominal accretion efficiency for the untilted (10h) and tilted (110h) high-resolution simulations.  The {\em top} panel shows the mass flux, $\dot{M}$ in arbitrary units.  The {\em middle} panel shows the specific angular momentum flux, with a comparison to \citet{Novikov73}.  The {\em bottom} panel shows the nominal efficiency, $1-\dot{E}/\dot{M}$, again with comparison to the predictions of NT73.}
\label{fig:fluxes}
\end{center}
\end{figure}

Time histories of the mass and angular momentum fluxes at the event horizon are shown in Figure \ref{fig:fluxes}.  The {\em top} panel shows the mass accretion rate, $\dot{M}(r_H,t)$, while the {\em middle} panel shows the flux of specific angular momentum at the horizon
\begin{equation}
\frac{\dot{J}(r_H,t)}{\dot{M}(r_H,t)} = \frac{-\int \sqrt{-g} T^r_\varphi d\vartheta d\varphi}{-\int \sqrt{-g} \rho u^r d\vartheta d\varphi}~.
\end{equation}
This value can be compared to the prediction of NT73, which is also shown in Figure \ref{fig:fluxes}.  We find that our value is 13\% less than the prediction.  Since NT73 assumes that internal stresses in the disk to vanish at the ISCO, the angular momentum flux through the event horizon should be equal to its value at the ISCO.  However, as Figure \ref{fig:profiles} shows, the specific angular momentum in our simulations continues to fall inside the ISCO, suggesting that stresses are still acting on the gas.  Thus, not as much angular momentum reaches the black hole.  The {\em bottom} panel of Figure \ref{fig:fluxes} includes the ``nominal'' efficiency, $1-\dot{E}/\dot{M}$, which represents the total loss of specific energy into the black hole, where
\begin{equation}
\dot{E}(t) = -\int \sqrt{-g} T^r_t d\vartheta d\varphi~.
\end{equation}
Here the time histories of both simulations are consistent with the results of other comparable simulations \citep[for example, see Figure 7 of][]{Penna10} and the predictions of NT73.  In particular, the nominal efficiency for a thin disk around a slowly rotating black hole is expected to be $\approx 0.06$.  Our simulations give values slightly higher than this, consistent with additional energy being liberated from the accreting gas after it passes the ISCO.  
The discrepancies in $\dot{J}/\dot{M}$ and $1-\dot{E}/\dot{M}$ are comparable to those reported in \citet{Noble10} and \citet{Penna10} for similar disk thicknesses and field topologies.  Again, the discrepancies would likely be smaller had we used smaller, alternating poloidal magnetic field loops in our initialization.

Noting that the mass accretion rate increases slightly inside the ISCO, in contradiction with the assumption of NT73, one could instead compare the angular momentum flux and nominal efficiency using the ISCO value of $\dot{M}$.  For model 10h, the mass accretion rate at the ISCO is 96\% of that at the event horizon (0.708 vs. 0.734). Using the ISCO value, instead of the event horizon one, would bring the $\dot{J}/\dot{M}$ data up slightly, but it would still lie below the predictions of NT73. For the nominal efficiency, changing $\dot{M}$ in this way would actually drop the simulation results below the NT73 prediction (to about 0.035). However, it is obviously inconsistent to combine different fluxes from different radii in this way; we merely mean to illustrate the sensitivity of these results to the exact fluxes. As for the cause of the increase in $\dot{M}$ inside the ISCO, this is not a systematic effect, but rather reflects time variations in the mass accretion rate that are not completely smoothed out by time averaging that we use.

\begin{figure}
\begin{center}
\includegraphics[width=0.9 \columnwidth,angle=0]{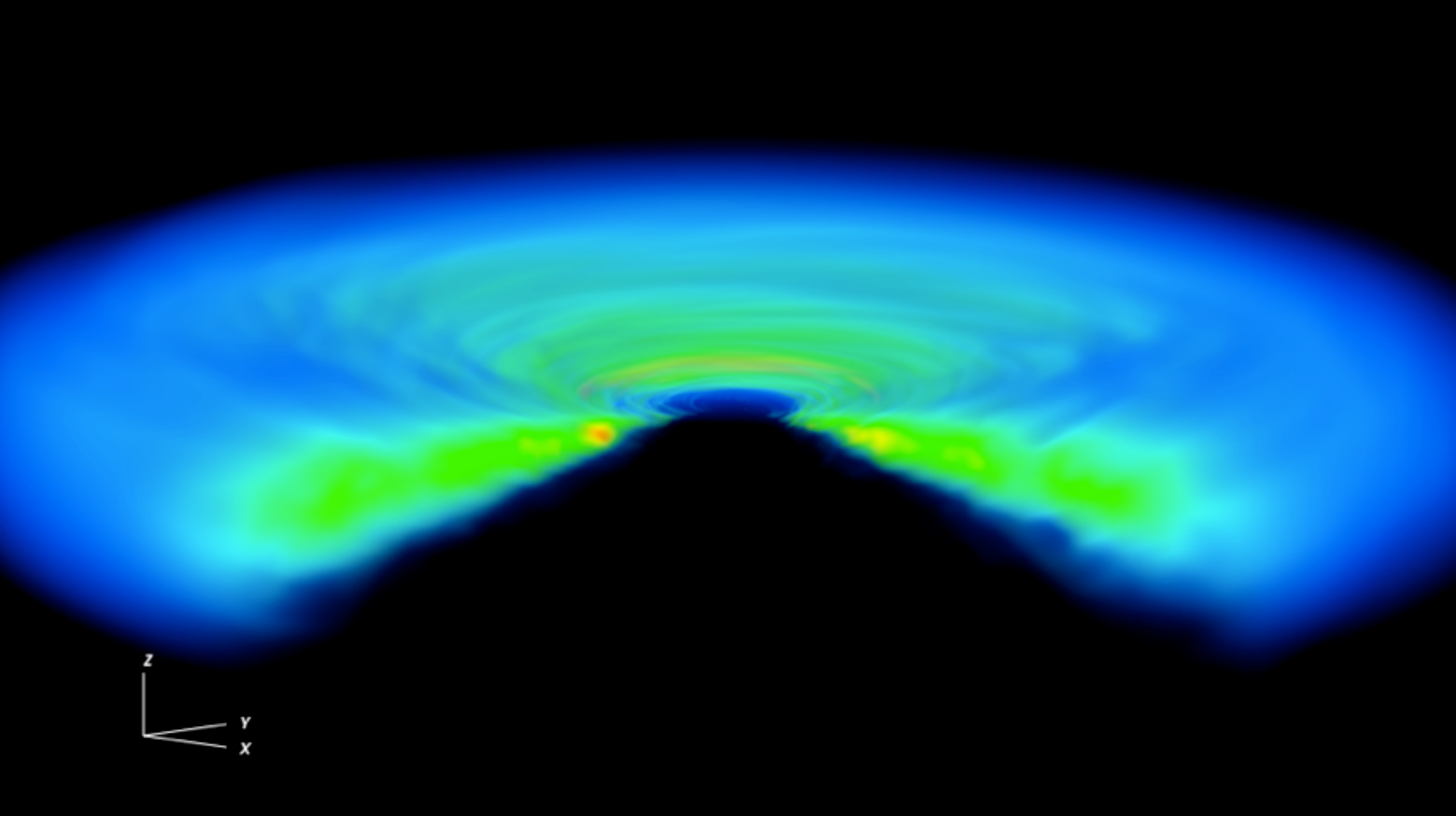}
\caption{Volume visualization of the logarithm of density (scaled from 0.01 to 1) at $t = 12.5 t_\mathrm{orb}$ for model 10h. A quarter of the disk has been cut away to reveal the cross section. The black hole spin axis is oriented vertically along the $z$-axis.}
\label{fig:10himage}
\end{center}
\end{figure}

Finally, in Figure \ref{fig:10himage}, we provide a volume-visualization of the end state of our high-resolution, untilted simulation 10h.  This will be useful for making a basic qualitative comparison with our tilted disk results, presented in the next section.  

\section{Results of Tilted Disk Simulations}
\label{sec:tilted}

\begin{figure}
\begin{center}
\includegraphics[width=0.9 \columnwidth,angle=0]{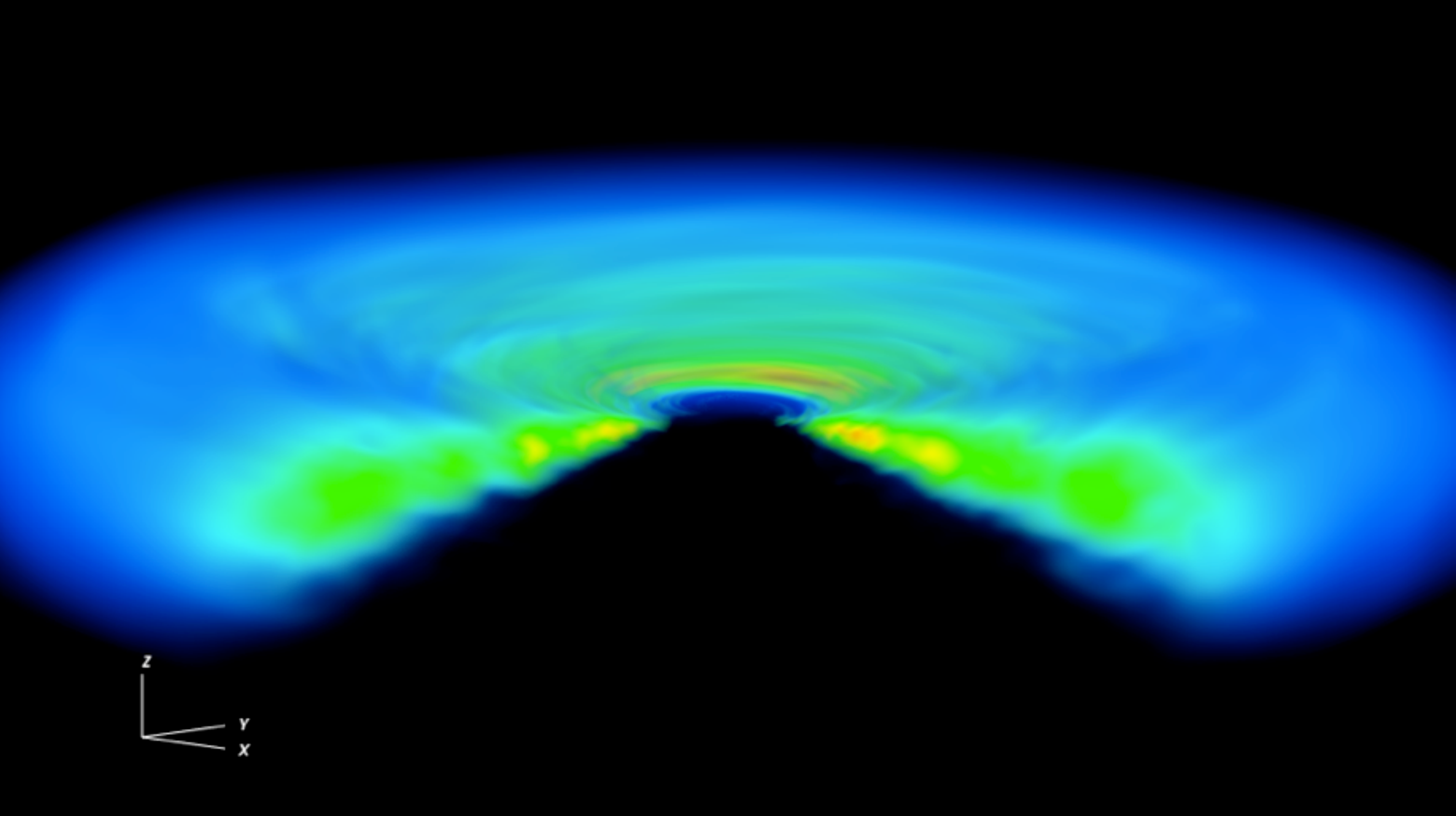}
\caption{Same as Fig. \ref{fig:10himage} except for model 110h.  In this case the black hole spin axis is oriented in the $x-z$ plane, tilted $10^\circ$ toward the $-x$-axis from the $+z$-axis.  No indication of warping appears to be present.  (An animation of this figure is available in the online journal.)}
\label{fig:110himage}
\end{center}
\end{figure}

Figure \ref{fig:110himage} shows a volume visualization of our high-resolution tilted simulation 110h at the same evolution time as Figure \ref{fig:10himage}.  The most remarkable thing to note is how similar the disks appear in both cases, giving the first indication that, even for these much thinner accretion disks (as compared to the ones in \citet{Fragile07}), there is no sign of Bardeen-Petterson alignment in the prograde case.  This is despite the fact that Figure \ref{fig:alpha} shows that for most of the time-resolved part of the disk ($r<12M$; see Figure \ref{fig:timescales}), $\overline{\alpha_{\hat{r}\hat{\phi}}} \gtrsim \langle\delta\rangle$, such that one might, conventionally, expect some Bardeen-Petterson-like behavior.  

\begin{figure}
\begin{center}
\includegraphics[width=0.9 \columnwidth,angle=0]{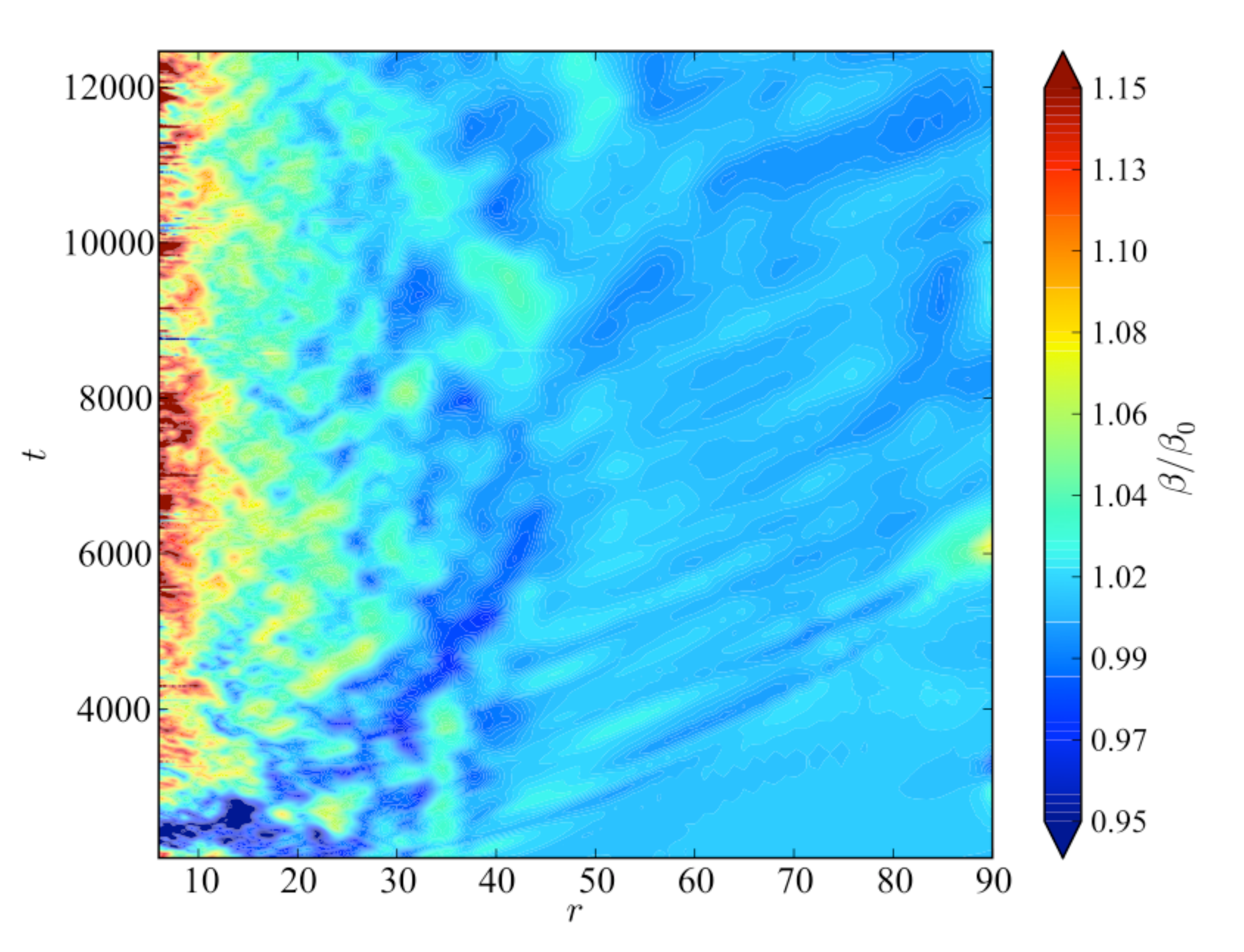}
\caption{Fractional disk tilt, i.e. $\beta/\beta_0$, as a function of radius and time for simulation 110h.  The plot shows no evidence for Bardeen-Petterson alignment at any radius over the duration of the simulation.  In fact, the inner disk tilts {\em away} from the black hole symmetry plane.}
\label{fig:beta}
\end{center}
\end{figure}

A more quantitative illustration of this statement is given in Figure \ref{fig:beta}, which plots the tilt, $\beta$, of model 110h as a function of radius for all times during the simulation (see Paper 2 for details on how we extract $\beta$ from the simulations).  Rather than alignment of the disk toward the black hole symmetry plane, this figure shows a tendency for $\beta$ to evolve away from zero, at least at small radii.  Furthermore, no part of the disk shows more than a 5\% alignment at any time.  The same results hold true for model 110m, which was run to a much later stop time.  The clear implication is that Lense-Thirring precession does not work to align the disk in this case.

\begin{figure}
\begin{center}
\includegraphics[width=0.9 \columnwidth,angle=0]{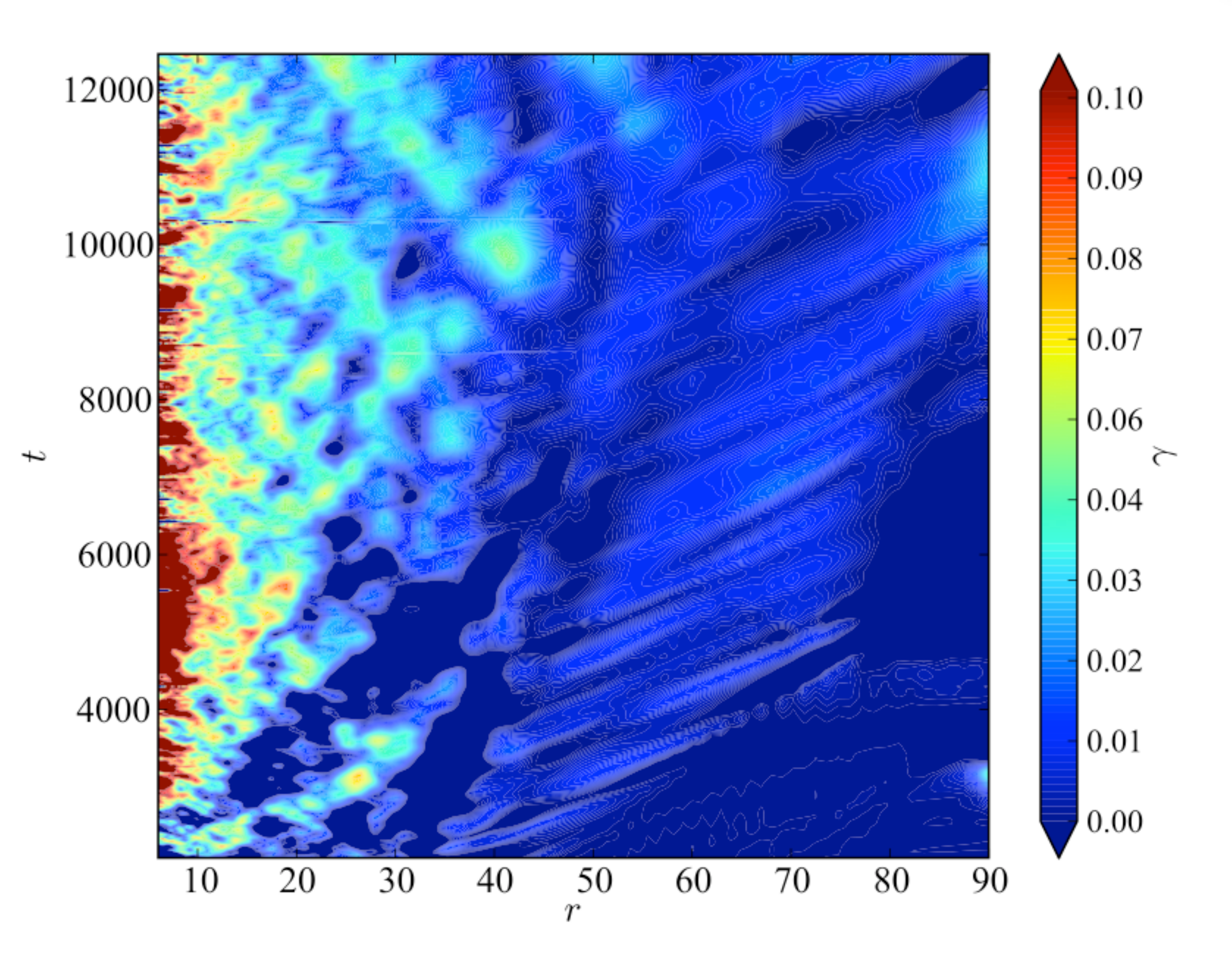}
\caption{Twist, $\gamma$, of each radial shell as a function of time for simulation 110h.  A ``twist front'' can be seen moving slowly out through the disk.  Behind this front, the twist seems to saturate at a certain value in each radial shell, such that only moderate differential precession is seen.}
\label{fig:gammaR}
\end{center}
\end{figure}

However, Lense-Thirring precession does still cause a twisting of the disk, although again, not in the simple manner that might be expected.  Figure \ref{fig:gammaR} shows a spacetime plot for the disk twist, $\gamma$, similar to the previous figure for tilt.  This plot reveals two important behaviors in the disk:  First, at small radii ($r < 10M$), after an initial period of strong differential precession, the twist saturates at a modest value and later relaxes toward a smaller one.  Meanwhile, the twist gradually builds up at larger radii, as the rest of the disk ``catches up'' with the twist of the inner disk.  By the end of simulation 110h, the twist front has moved out to about $40M$.  A roughly similar pattern of strong initial differential precession, followed by more gradual global precession was seen in \citet{Sorathia13}.  Generally, somewhat stronger differential precession has been seen in SPH simulations \citep[e.g.][]{Nelson00}, although direct comparisons are difficult as the parameters of SPH and GRMHD simulations are often quite different.

\begin{figure}
\begin{center}
\includegraphics[width=0.9 \columnwidth,angle=0]{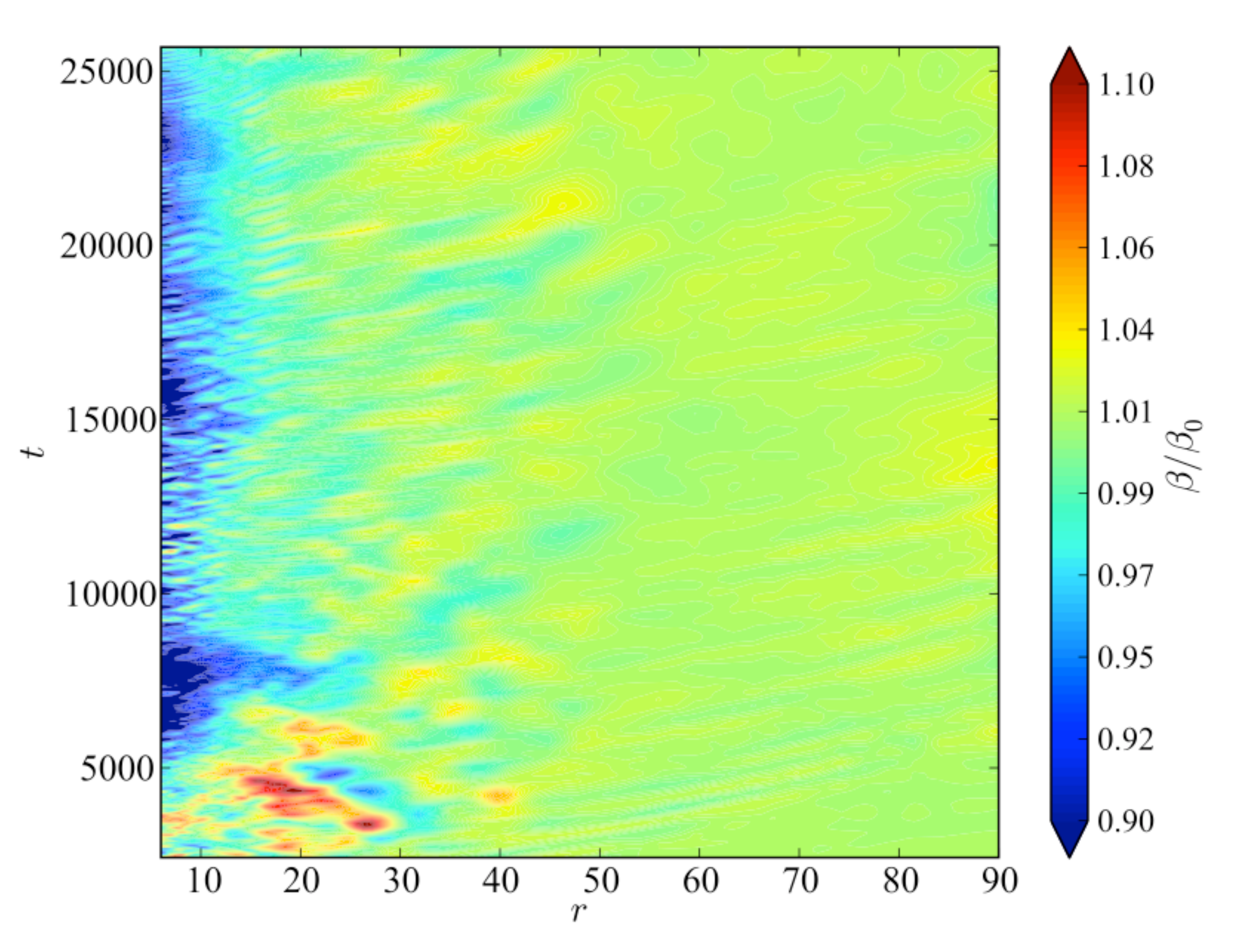}
\caption{Same as Figure \ref{fig:beta} except for simulation 110rm.  Here $\beta$ is measured as the tilt away from anti-alignment (so a value of 0 would be fully anti-aligned).  There is modest evidence for evolution toward (anti-) alignment at small radii ($r \lesssim 10 M$).  This holds true for most of the duration of the simulation ($5000 < t < 26000$).}
\label{fig:beta_retro}
\end{center}
\end{figure}

Unlike the prograde cases, the retrograde case, 110rm, does exhibit some tendency to (counter-) align, as shown in Figure \ref{fig:beta_retro}, although the effect is still not strong.  The alignment inside of $r = 10M$ is only about 10\%, and this remains true for most of the duration of the simulation.  Although this is not a large value, it is consistent with the predictions of our semi-analytic model, as we show in Paper 2.  Using those results as a guide, we predict that thinner disks should exhibit even greater alignment, with full alignment at the inner edge of the disk ($\beta(r_\mathrm{ISCO}) < 0.1 \beta_0$) likely to occur when $\delta \lesssim 0.04$, for $\alpha = 0.1$ and $a_*=-0.1$.  Thus, complete alignment of a disk plane with a black hole symmetry plane may be more likely to occur for retrograde systems than for prograde ones.  

All of these results are consistent with the linear analytic theory of tilted disks whenever the effective viscosity is small, $\alpha < \delta$ \citep[see e.g.][]{Ivanov97,Nelson00,Lubow02,Zhuravlev11}. In particular, in this limit, analytic theory predicts alignment is only possible whenever the directions of nodal and apsidal precessions are opposite to each other, while whenever they are directed in the same sense, either growth of the disk inclination angle toward the black hole or  radial oscillations of this angle are predicted.  Furthermore, in Paper 2 we show that all these results are qualitatively consistent with the predictions of our semi-analytic model, although there are some quantitative discrepancies.  We also differentiate which physical effects in the disk are most responsible for our results.

\subsection{Standing Shocks}
\label{sec:shocks}

\begin{figure}
\begin{center}
\includegraphics[width=0.9 \columnwidth,angle=0]{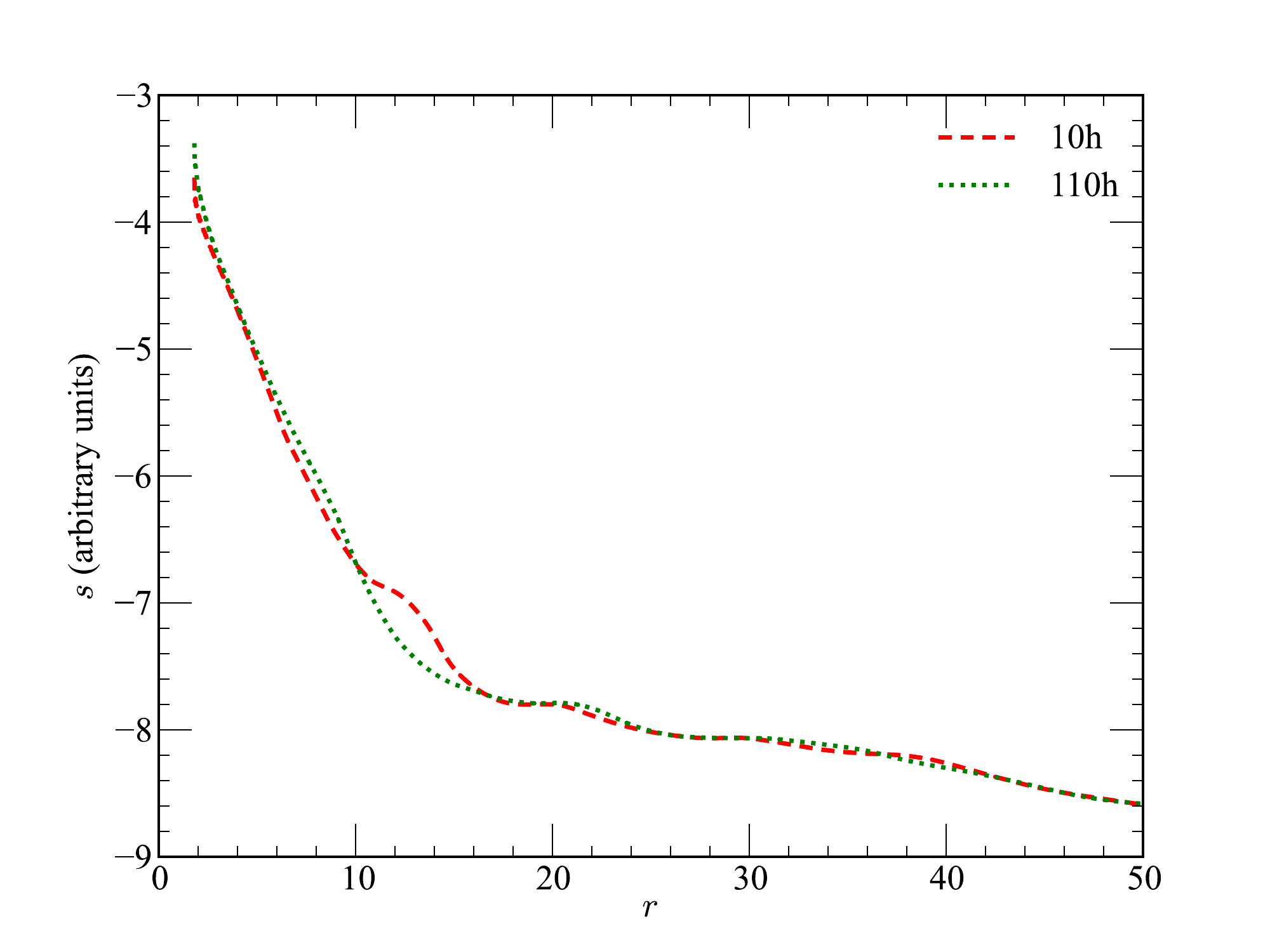}
\caption{Density-weighted radial shell averages of the entropy, $s$, (arbitrary units) for the untilted (10h) and tilted (110h) high-resolution simulations.  Data are time averaged over the interval $10 \le t/t_\mathrm{orb} \le 12.5$.  The similarity of both profiles suggests that standing shocks do not play an important role in the tilted simulation.}
\label{fig:entropy}
\end{center}
\end{figure}

One important defining characteristic of the tilted accretion disks in our previous numerical studies was the presence of standing shocks at small radii \citep{Fragile08}.  These shocks were found to be responsible for a number of unique phenomena.  They lead to a larger disk truncation radius in simulated tilted disks \citep{Fragile09b,Dexter11}; they may help amplify the inherent variability of accretion disks \citep{Henisey12}; and they dissipate a significant fraction of the accreted rest mass energy \citep{Generozov14}.  However, we find no evidence for similar shocks in the simulations we present in this paper.  One measure of this is to look at the density-weighted shell average of entropy, $s\equiv \log(P_g/\rho^\Gamma)$, as in Figure \ref{fig:entropy}.  Since shocks generate entropy, the association of shocks with tilted simulations should manifest itself through an excess of entropy \citep[see, e.g., Figure 14 of][]{Dexter11}.  However, such an excess is not seen in Figure \ref{fig:entropy}; instead the entropy of the untilted and tilted simulations track each other quite closely at all radii.  We also point out that the general rise in entropy toward smaller radii, starting from around $r = 14M$, is actually associated with the gradual decline of disk density (see Figure \ref{fig:profiles}) and dissipation of magnetic field energy toward the black hole. Other signatures of standing shocks, such as regions of post-shock density enhancement, are also missing in simulation 110h.

\begin{figure}
\begin{center}
\includegraphics[width=0.9 \columnwidth,angle=0]{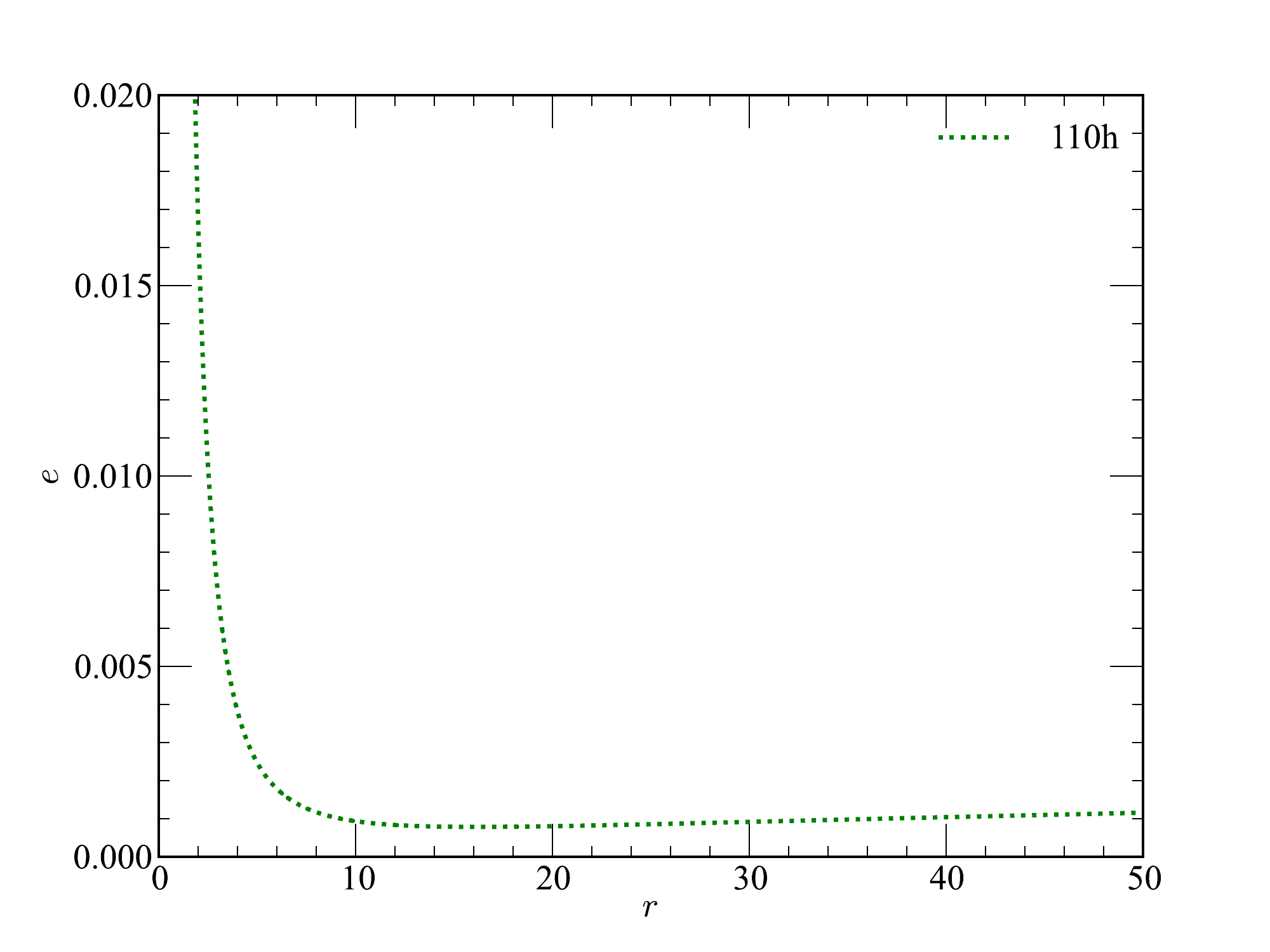}
\caption{Shell-averaged orbital eccentricity, $e$, for simulation 110h, estimated at one scale height in the disk. Data are time averaged over the interval $10 \le t/t_\mathrm{orb} \le 12.5$.}
\label{fig:eccentricity}
\end{center}
\end{figure}

The physical explanation put forward in \citet{Fragile08} for the standing shocks is that they are a response of the disk to the crowding of particle trajectories near their apocenters whenever the eccentricity of those trajectories increases significantly toward smaller radii \citep{Ivanov97}.  The orbital eccentricity, itself, is a manifestation of the epicyclic driving of the gas due to unbalanced radial pressure gradients found at high latitudes in tilted disks.  To quantify these statements, we can use the criterion proposed in \citet{Ivanov97} that shocks will form whenever the eccentricity, $e$, becomes comparable to $\delta$.  This is based upon the assumption that a typical value for the velocity perturbations associated with the eccentricity is $v^{'}\sim er\Omega$ and that the sound speed is $~H\Omega$. For tilted, warped disks, we can estimate the eccentricity at one scale height in the disk as 
\begin{equation}
e (H=r) = -  \frac{r}{6M} \frac{\partial(\beta \sin \gamma)}{\partial r} ~.
\end{equation}
In order to evaluate this expression, we first fit the shell-averaged $\beta$ and $\gamma$ with power laws \citep[for this part only, we use the $\beta$ and $\gamma$ obtained from Equations (32) and (41), respectively, of][]{Fragile07}.  The result for simulation 110h is shown in Figure \ref{fig:eccentricity}.  Unlike our previous results \citep[see Figure 16 of][]{Dexter11}, the eccentricity in this simulation is very small and the inequality $e \ll \delta$ is satisfied by a wide margin. Thus, the lack of shocks in this simulation is perfectly consistent with our understanding of what generates the standing shocks in our other tilted simulations.  In this particular case, it seems the combination of small black hole spin, small tilt, and small scale height prevent the standing shocks from forming.

\subsection{Timescales}
\label{sec:timescales}

\begin{figure}
\begin{center}
\includegraphics[width=0.9 \columnwidth]{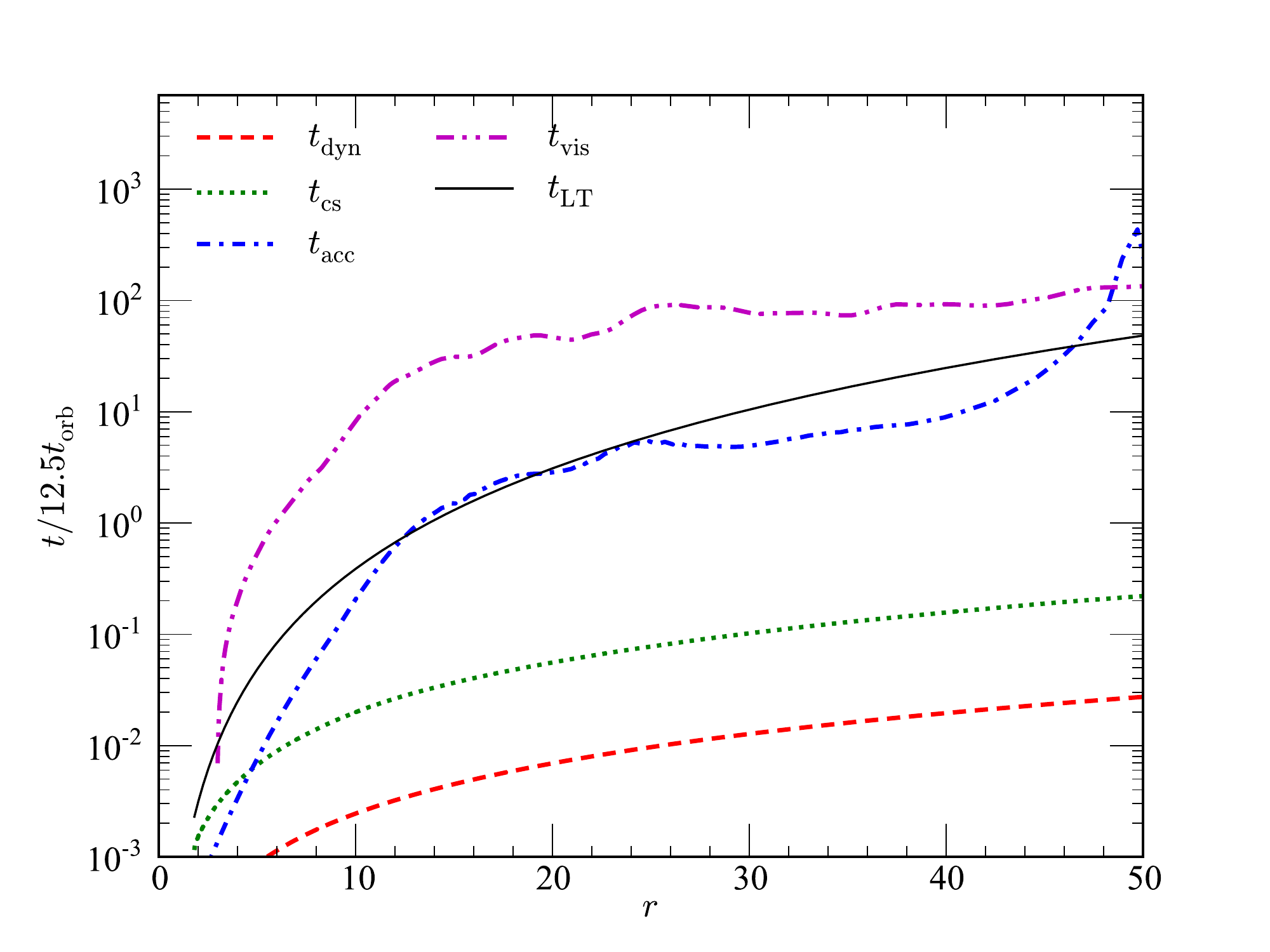}
\caption{The dynamical, sound-crossing, accretion, viscous, and Lense-Thirring precession timescales from our tilted, high-resolution simulation (110h), time-averaged over the interval $10 \le t/t_\mathrm{orb} \le 12.5$.  All values are normalized by $t = 12.5 t_\mathrm{orb} = 13000 M$, the duration of this simulation.  By doing so we can easily see which timescales are covered, and over what radial range, during the course of this simulation.}
\label{fig:timescales}
\end{center}
\end{figure}

Another way to think of tilted disks is in terms of timescales.  For tilted black hole accretion disks, there are a number of relevant ones to consider: 1) the dynamical timescale, $t_\mathrm{dyn} = 1/\Omega$; 2) the radial sound-crossing time, $t_\mathrm{cs} = r/c_s$; 3) the accretion timescale, $t_\mathrm{acc} = r/V^r$; 4) the viscous timescale, $t_\mathrm{vis} = r^2/\alpha_{\hat{r}\hat{\phi}} c_s H$; and 5) the Lense-Thirring precession time, $t_\mathrm{LT} = 1/\Omega_\mathrm{LT}$, where $\Omega_\mathrm{LT}=2aM/r^3$ is the Lense-Thirring precession frequency.  Detailed explanations of these and other timescales are provided in Paper 2.  Figure \ref{fig:timescales} shows a plot of these timescales as a function of radius for simulation 110h. Other than $t_\mathrm{LT}$, each of the timescales is generated from density-weighted averages of the relevant fluid variables, $\Omega$, $c_s$, $V^r$, $\alpha_{\hat{r}\hat{\phi}}$, and $H$.  From this figure, we see that the duration of our high-resolution simulation 110h, $t = 13000 M = 12.5 t_\mathrm{orb}$, is enough to cover: the viscous timescales out to $r=6 M$; the Lense-Thirring and accretion timescales out to $r=13 M$; and the two remaining timescales at all radii.

According to the original Bardeen-Petterson picture, a tilted disk should align with the black hole wherever $t_\mathrm{LT} < t_\mathrm{vis}$, which is most likely to occur close to the black hole, where $t_\mathrm{LT}$ is shortest.  In Figure \ref{fig:timescales}, we see that this is the case all the way out beyond $r > 50M$, though at such large radii both timescales are significantly longer than the duration of our high-resolution simulation, so caution should be used when considering those data.  The fact that $t_\mathrm{LT} < t_\mathrm{vis}$ over much of the disk, would seem to suggest the disk should align with the black hole.  However, as mentioned in Section \ref{sec:intro}, the Bardeen-Petterson argument is known to be quantitatively incorrect. In Paper 2 we introduce a relaxation radius and timescale that provide a better explanation for what is going on here.

The behavior of $\gamma$ in our simulations can be understood in terms of the sound-crossing time, $t_\mathrm{cs}$.  Based on our previous work \citep{Fragile05,Fragile07}, we have argued that if the sound speed in tilted disks is high enough, or equivalently the sound-crossing time short enough, then the Lense-Thirring torque rapidly redistributes itself throughout the body of the disk.\footnote{This same effect has been discussed in the context of non-relativistic twisted disks in binary systems \citep[see e.g.][]{Papaloizou95, Larwood96}.}  The result is that the disk as a whole experiences rigid-body precession.  As we commented in Section \ref{sec:tilted}, the spacetime diagram of $\gamma$ in Figure \ref{fig:gammaR} shows that, after an initial period of differential precession, the radial profile of $\gamma$ does not change significantly at small radii, while the precession continues to grow gradually at larger radii.  Our previous results suggest that once the rest of the disk starts precessing, the whole disk will precess together \citep[e.g.][]{Fragile08}.

\begin{figure}
\begin{center}
\includegraphics[width=0.9 \columnwidth,angle=0]{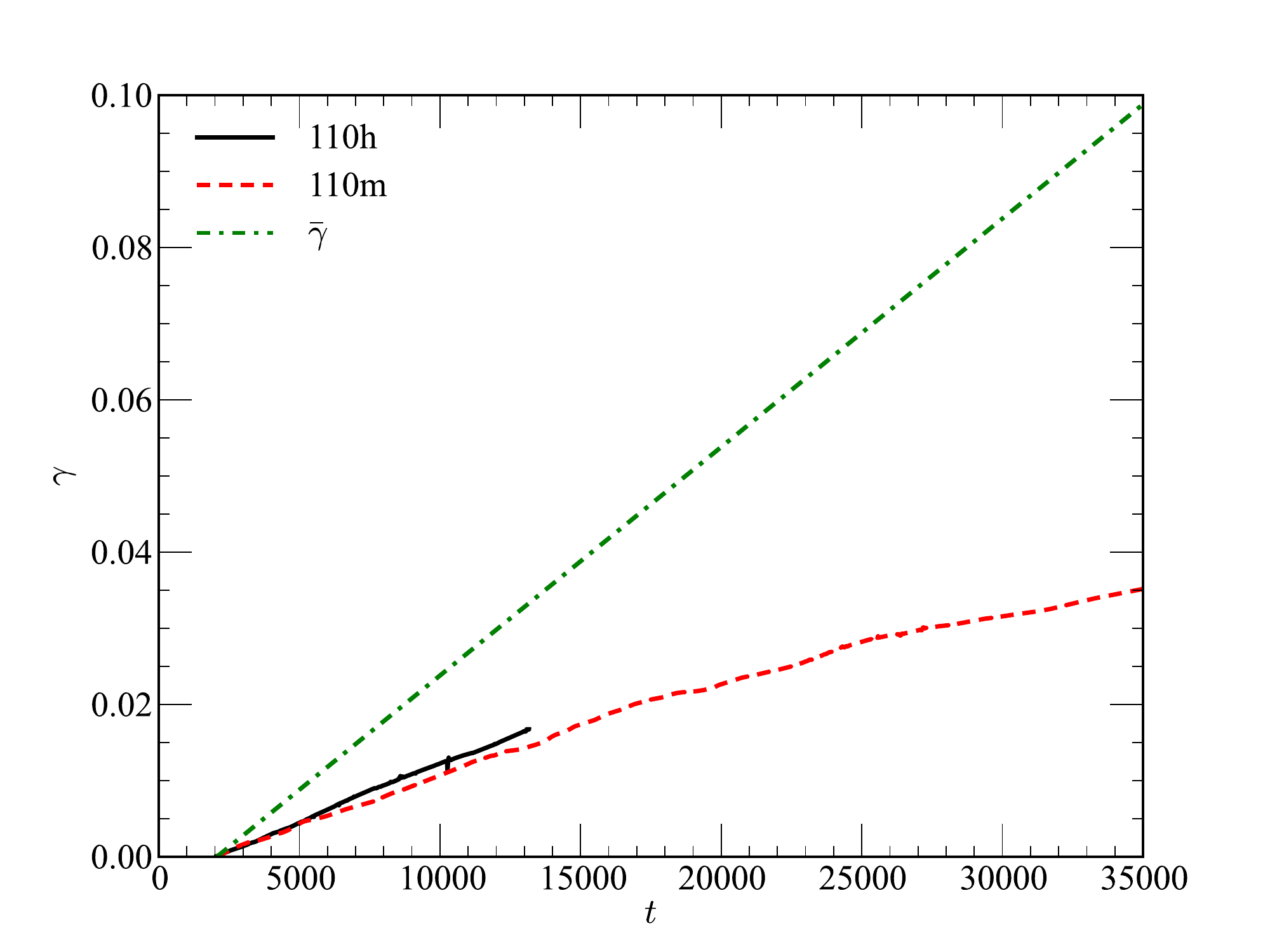}
\caption{Precession (or twist) angle, $\gamma$, averaged over the bulk of the disk ($5 \le r/M \le 90$), as a function of the time for simulations 110h and 110m.  We also plot the simple estimate of $\overline{\gamma}$ from equation (\ref{ennn}).}
\label{fig:gammaT}
\end{center}
\end{figure}

This motivates us to define a total angular momentum vector for the disk (defined over the region $5 \le r/M \le 90$), and see if we find evidence of this vector precessing.  The result is shown in Figure \ref{fig:gammaT}, where we plot the instantaneous precession angle of the disk angular momentum vector as a function of time.  For comparison, we can also estimate the precession timescale, $t_\mathrm{prec}$, for the disk (see Paper 2 for details).  For small values of the average twist, $\overline \gamma$, when the disk is close to $\overline \beta \approx \beta_0$, this timescale is
\begin{equation}
t_\mathrm{prec} \approx \frac{\int r^{3/2} dr \Sigma}{2a_* \int r^{-3/2} dr \Sigma} ~,
\end{equation}
where we have assumed solid-body precession and ignored relativistic corrections and corrections proportional to $\cos \beta_0$.  Since the surface density is a function of time, this precession timescale is also formally time dependent.  However, we have checked that for all times available in our high resolution simulations, the value is close to $t_\mathrm{prec}\approx 3.4 \times 10^5 M$. From this we get that the average precession angle should depend on time as 
\begin{equation}
\overline \gamma = \frac{t}{t_\mathrm{prec}} \approx 3 \times 10^{-6}t ~.
\label{ennn}
\end{equation}
This prediction is plotted against the simulation data in Figure \ref{fig:gammaT}.  Although the agreement between the predicted and measured slopes is not great, it is within the implied (order unity) uncertainties in equation (\ref{ennn}). These uncertainties come from the fact that the full calculation of $t_\mathrm{prec}$ involves evaluating a fraction where the numerator is the difference of two large, nearly equal numbers, and the denominator is small (again, see Paper 2 for details).  This type of operation is prone to the accumulation of numerical errors.  Furthermore, these errors are sensitive to our choice of a small black hole spin and small tilt. To confirm that our method is, nevertheless, sound, we performed a test on a purely hydrodynamic simulation of a tilted disk with a much larger black hole spin ($a_* = 0.9$) and found that the measured value for precession matched our analytic prediction to better than 1\%. 

\section{Turbulent Stress Tensor}
\label{sec:viscosity}

Much of the analytic, and even numerical SPH, work on tilted disks relies in some way on the validity of the Boussinesq approximation, where an effective stress tensor describing the action of turbulent motions on the mean flow is postulated to be proportional to the sum of an isotropic tensor, giving an additional contribution to the pressure term, and the rate-of-strain tensor. The latter part describes dissipation of the energy of the mean flow and is proportional to an effective viscosity coefficient, which is, in turn, proportional to the Shakura-Sunyaev parameter $\alpha$. The Boussinesq approximation is based on the assumption that turbulent motions are isotropic in a statistical sense. However, recent work has called this assumption into question \citep{Sorathia13,Nauman14}.  

We test the isotropy of the turbulence here for the first time using GRMHD simulations.  To do so, we calculate both the stress ($W_{\hat{i}\hat{j}}$) and rate-of-strain ($S_{\hat{i}\hat{j}}$) tensors in the co-moving frame of the fluid (see Appendix \ref{app:viscosity} for how we do this).  If the stress tensor were indeed isotropic, then, as we mentioned above, it should be proportional to a multiple of the rate-of-strain tensor plus an isotropic tensor.  We would, therefore, expect the ratio between the off-diagonal components of the stress tensor and the same off-diagonal components of the rate-of-strain tensor ($W_{\hat{i}\hat{j}}/S_{\hat{i}\hat{j}}$) to be a constant, which we could use to define an ``alpha'' for our disk ($\alpha_{\hat{i}\hat{j}} = -3W_{\hat{i}\hat{j}}\Omega/(4S_{\hat{i}\hat{j}} P_\mathrm{Tot})$, where $P_\mathrm{Tot} = P_g + P_B$).  This is probably too strong a statement, as the turbulence is more likely to be {\em nearly} isotropic, rather than exactly so.  In this case, it might be sufficient for us to find that all the $\alpha_{\hat{i}\hat{j}}$ have a consistent sign and similar magnitude.  Instead, we find that, while $\alpha_{\hat{r}\hat{\phi}}$ behaves as expected for thin accretion disks (with a consistent, positive sign and magnitude in the range 0.01-0.1 inside the disk), the other components, $\alpha_{\hat{r}\hat{\theta}}$ and $\alpha_{\hat{\theta}\hat{\phi}}$, vary wildly in magnitude, and are not even consistent in sign.  

\begin{figure}
\begin{center}
\includegraphics[width=0.9 \columnwidth,angle=0]{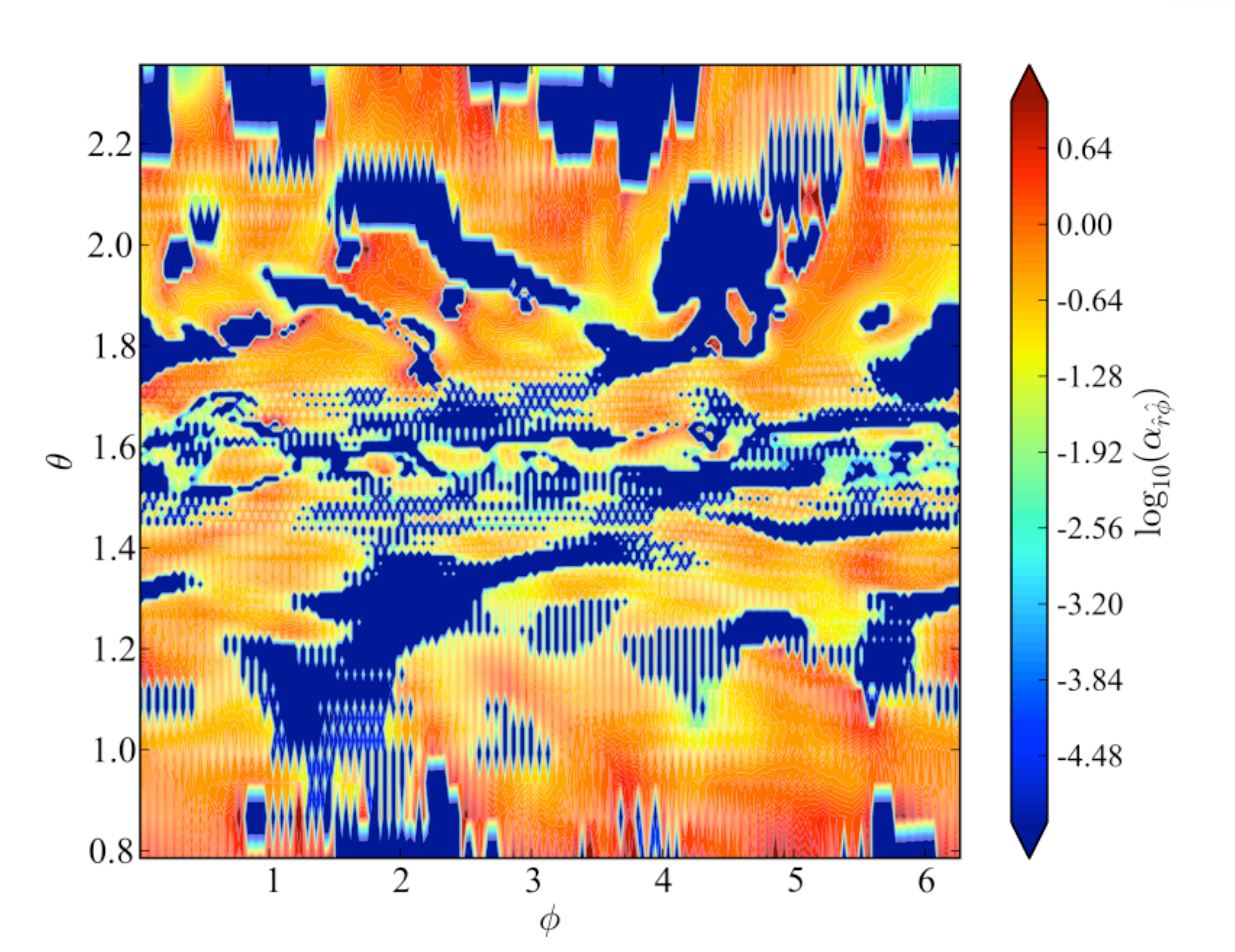}
\caption{Pseudocolor plot of $\log_{10}(\alpha_{\hat{r}\hat{\phi}})$ on the $r=10 M$ shell at the end of the 110h simulation ($t = 12.5 t_\mathrm{orb} = 13000 M$).  Most of this shell exhibits a consistent sign for $\alpha_{\hat{r}\hat{\phi}}$ and a magnitude between $0.01-1$.}
\label{fig:alphaRP}
\end{center}
\end{figure}

\begin{figure}
\begin{center}
\includegraphics[width=0.8 \columnwidth,angle=0]{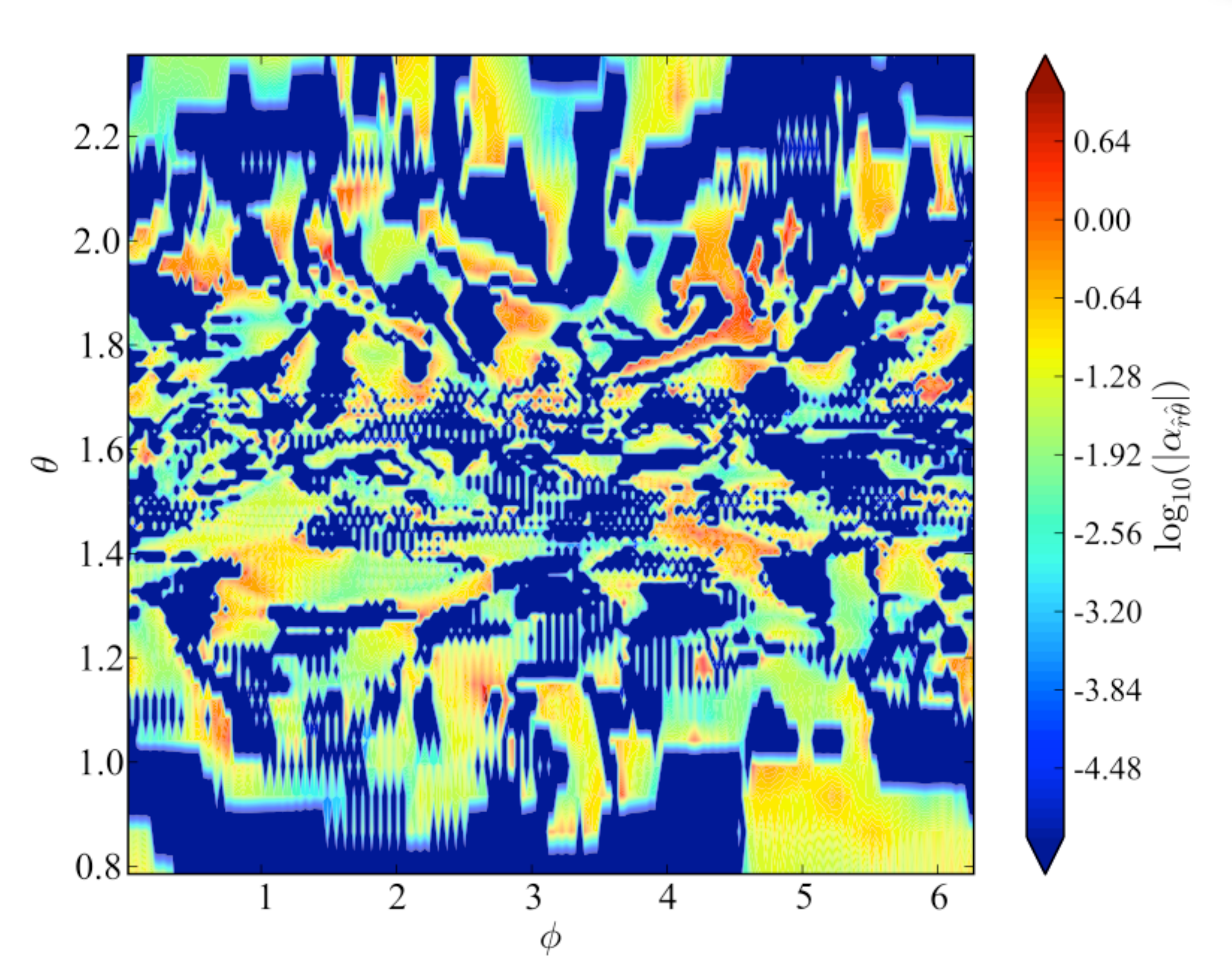}
\includegraphics[width=0.8 \columnwidth,angle=0]{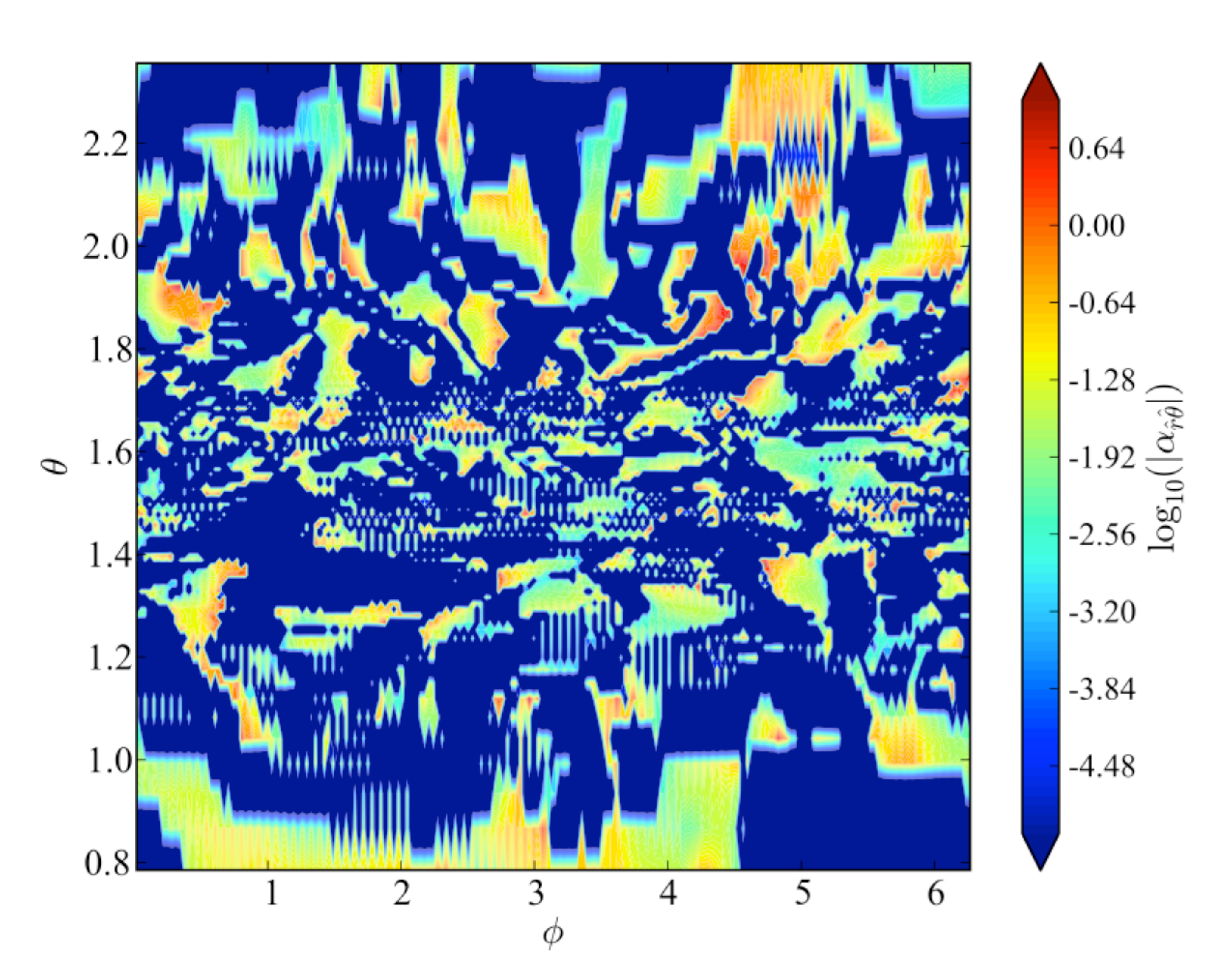}
\caption{Pseudocolor plot of $\log_{10}(\vert\alpha_{\hat{r}\hat{\theta}}\vert)$ on the same shell and at the same time as Figure \ref{fig:alphaRP}.  Since $\alpha_{\hat{r}\hat{\theta}}$ does not have a consistent sign, in order to plot $\log_{10}$ values, we had to split the plot into two panels.  The {\em top} panel is for when $\alpha_{\hat{r}\hat{\theta}} > 0$ and the {\em bottom} panel is for when $\alpha_{\hat{r}\hat{\theta}} < 0$.  This figure demonstrates that $\alpha_{\hat{r}\hat{\theta}}$ alternates in sign more frequently and covers a much broader range of magnitudes than $\alpha_{\hat{r}\hat{\phi}}$.}
\label{fig:alphaRT}
\end{center}
\end{figure}

For example, Figures \ref{fig:alphaRP} and \ref{fig:alphaRT} show how $\alpha_{\hat{r}\hat{\phi}}$ and  $\alpha_{\hat{r}\hat{\theta}}$ vary, respectively, with spatial position over the $r=10 M$ radial shell at the end of simulation 110h.  A plot of $\alpha_{\hat{\theta}\hat{\phi}}$ is not included, though looks very similar to that of $\alpha_{\hat{r}\hat{\theta}}$.  One sees in Figure \ref{fig:alphaRP} that $\alpha_{\hat{r}\hat{\phi}}$ is nearly everywhere consistent in sign, with magnitudes between 0.01 and 1 over most of the shell.  This plot is from a single time slice, so, as expected, there is some spatial variability, even in the azimuthal direction.  Furthermore, none of the components are shell-averaged in this figure, as they were in Figure \ref{fig:alpha}.  The two panels of Figure \ref{fig:alphaRT}, on the other hand, look very different from Figure \ref{fig:alphaRP}.  The very fact that we have to show two panels, one to represent positive values of $\alpha_{\hat{r}\hat{\theta}}$ and one to represent negative, indicates that, unlike $\alpha_{\hat{r}\hat{\phi}}$, the signs of the other components of $\alpha_{\hat{i}\hat{j}}$ are almost equally likely to be negative as positive.  We also see that the magnitudes vary over a larger range, $\sim 10^{-3} - 1$.

These results are broadly consistent with those presented in \citet{Sorathia13}, the only other paper of which we are aware that reported on the isotropy of the stress tensor in global MHD simulations.  Note that their $\alpha_*$ only considers the $r-\theta$ component of the Maxwell stress, and does not include the Reynolds stress.  It generally has a magnitude in the range $\sim 10^{-5} - 10^{-4}$, while they state elsewhere in their paper that the Reynolds stress for this component can often approach $\sim 1-10 P_g$.  Since our $\alpha_{\hat{r}\hat{\theta}}$ includes both the Maxwell and Reynolds contributions, it seems our results are consistent in both a qualitative and quantitative sense with theirs.  About the only difference is that the figures in \citet{Sorathia13} show much finer spatial structure, as would be expected given their higher resolution.

There are several possible explanations for the apparent anisotropy of the turbulence:  First, even assuming that there is a local-in-time correlation between the stress and rate-of-shear tensors, the relationship between them may not be a direct proportionality of their respective components.  In an anisotropic medium, a general linear relation could be of the form $W_{\hat{i}\hat{j}}\sim D_{\hat{i}\hat{j}\hat{l}\hat{m}}S_{\hat{l}\hat{m}}$, where the tensor $D_{\hat{i}\hat{j}\hat{l}\hat{m}}$ is not necessarily diagonal.  Secondly, there could be non-local-in-time correlations between $W_{\hat{i}\hat{j}}$ and $S_{\hat{i}\hat{j}}$.  For example, \citet{Hirose09} found in shearing-box simulations that there could be a time lag between build-ups of $W_{\hat{i}\hat{j}}$ and pressure in the disk. This situation can either be described by expressing the relationship between  $W_{\hat{i}\hat{j}}$ and $S_{\hat{i}\hat{j}}$ in terms of a convolution with an appropriate kernel\footnote{Such convolutions have been discussed by, e.g. \citet{Ivanov04}, in the context of tidal forcing of a convective star.} or using a dynamical equation for $W_{\hat{i}\hat{j}}$ obtained from a suitable closure procedure \citep[see, e.g.][for a discussion]{Kato94,Ogilvie03}.

\section{Conclusions}
\label{sec:conclusion}

In this paper we have presented results of the thinnest tilted disk simulations yet done with a GRMHD code.  Performing simulations of thinner disks is important, as the behavior of tilted disks is expected to be qualitatively different for thin disks from thick ones.  The present simulations are roughly where we expect to begin to see this change of behavior.

Since these are some of the thinnest simulations we have thus far performed, we took this opportunity to compare our untilted reference simulations with earlier such simulations in the literature.  Despite our very modest resolution of the MRI, we generally find that our simulations reproduce the expected trends.  Furthermore, our numerical results approach the limit of the \citet{Novikov73} disk model over the region where we have achieved a reasonable inflow equilibrium.  Evidence of this include the relatively constant value of the specific angular momentum inside the ISCO, as well as an observed accretion efficiency close to the expected value.

The main focus of this work, however, is on tilted disk behavior.  Here we were particularly interested in whether or not our simulations would exhibit Bardeen-Petterson-like alignment, as the disk scale height was chosen to be comparable to the effective $\alpha$.  There are obviously many ways one could set up such a numerical experiment. Motivated by our own previous work, we chose to initialize the simulations with an isolated, tilted torus, threaded by weak poloidal magnetic fields. This triggered the formation of an MRI turbulent disk that we then cooled, as needed, to maintain the chosen scale height.  As with any numerical experiment, one hopes that the results are not strongly sensitive to the initial conditions, although this can not generally be proven without further numerical experiments.

For our prograde simulations, we did not observe Bardeen-Petterson alignment, instead finding that the tilt of the disk grows larger at small radii.  This type of behavior is more consistent with the bending wave behavior expected in thicker disks. We tentatively conclude that we are not yet in the parameter space of the Bardeen-Petterson solution. However, we can not be certain that the Bardeen-Petterson solution applies in MRI turbulent disks, as it relies on the assumption of an isotropic viscosity, which is not supported by our simulations. In contrast, we did observe some modest alignment of the disk in our retrograde simulation.  In this case, however, even bending wave theory predicts some alignment.  Overall, our results appear to be consistent with those analytic and semi-analytic models that fully account for all relativistic effects, a point we explore further in Paper 2.

In this paper, we saw some evidence for solid-body precession of the disk in each simulation, consistent with our earlier work.  Such behavior may be important for explaining some types of quasi-periodic oscillations (QPOs) seen in black hole X-ray binaries \citep{Ingram09, Motta14}.  However, this should be explored further, as there are significant unanswered questions about this behavior.  First, it is not clear how long a tilted disk can continue to precess.  In \citet{Fragile08}, we confirmed that an isolated thick disk can precess more or less as a solid body for at least one precession period, but to explain QPOs, disks would need to be able to sustain this behavior for very many precession periods.  The second question has to do with what influence a large outer disk, and even a binary companion, would have on this precession \citep[see][for a recent discussion]{Tremaine14}.  All of these are issues that should be explored further in future work.

Unlike our earlier work, we saw no evidence for standing shocks in the tilted simulations presented in this paper.  We attribute this to the small black hole spin, small tilt, and small disk scale height explored in this paper.  Together, these prevent the formation of significant, unbalanced radial pressure gradients in the disk.  Such radial pressure gradients are crucial for driving epicyclic motion, leading to non-circular orbits near the black hole.  If the eccentricity of these orbits increases toward smaller radii, then there is a crowding of orbits, leading to the formation of standing shocks.  However, we find that the eccentricity is very small, and nearly constant, in the thin, slightly tilted disks we explore in this paper, thus preventing the formation of such shocks.

Finally, we confirmed a point that has recently been highlighted by other authors \citep[e.g.][]{Sorathia13,Nauman14}, namely that the turbulent stresses in these simulated disks is definitely not isotropic.  At this stage, it is difficult to guess what implications this result has for the theory of tilted and warped accretion disks.  We explore this issue to some degree in Paper 2, although a detailed analysis must be postponed until a deeper theoretical understanding of this phenomenon is achieved.  

Our results are, of course, tentative until they can be confirmed by further numerical work.  Our main concerns are the relatively poor resolution and short evolution times of our simulations.  However, remedying these will require significantly greater amounts of computing time than were used here, and so will have to wait.

\acknowledgements
We thank Omer Blaes, Julian Krolik, Gordon Ogilvie, and the anonymous referee for their helpful feedback.  This work was supported in part by the National Science Foundation under Grant No. NSF PHY11-25915 and by NSF Cooperative Agreement Number EPS-0919440 that included computing time on the Clemson University Palmetto Cluster.  DMT was supported by a CAPES scholarship (proc. n$^\circ$ 1073-11-7) and thanks the College of Charleston, where part of this work was carried out, for their hospitality. VVZ and PBI were supported in part by Federal programme ``Scientific personnel'' contract 8422 and by programme 22 of the Presidium of Russian Academy of Sciences. Additionally, PBI was supported in part by RFBR grant 11-02-00244-a and the Grant of the President of the Russian Federation for Support of Leading Scientific Schools of the Russian Federation NSh-4235.2014.2 and VVZ was supported in part by grant RFBR-GFEN 14-02-91172 and in part by grant RFBF 12-02-00186.  The authors also acknowledge the Texas Advanced Computing Center (TACC) at The University of Texas at Austin for providing HPC resources that have contributed to the research results reported within this paper.  Finally, this work has made use of the computing facilities of the Laboratory of Astroinformatics (IAG/USP, NAT/Unicsul), whose purchase was made possible by the Brazilian agency FAPESP (grant 2009/54006-4) and the INCT-A.

\appendix

\section{Calculation of Turbulent Stress Tensor}
\label{app:viscosity}

The components of the stress and rate-of-strain tensors (in the co-moving frame) have the form $W_{\hat{i}\hat{j}}=e^\mu_{\hat{i}} e^\nu_{\hat{j}} T_{\mu\nu}^\mathrm{BL}$ and $S_{\hat{i}\hat{j}} = e^\mu_{\hat{i}} e^\nu_{\hat{j}} \sigma_{\mu\nu}^\mathrm{BL}$, where $e^\mu_{\hat{i}}$ are the basis vectors describing the local rest frame of the fluid.  Note that, to be consistent with previous estimates of the effective viscosities \citep[e.g.][]{Beckwith08,Penna10,Penna13}, we calculate the stress-energy and rate-of-strain tensors in Boyer-Lindquist coordinates.  Since our calculations are done in Kerr-Schild coordinates, we need to transform the stress energy (\ref{eq:tmunu}) as follows 
\begin{equation}
T_{\mu\nu}^\mathrm{BL}=\frac{\partial x^\alpha_\mathrm{[KS]}}{\partial x^\mu_\mathrm{[BL]}}\frac{\partial x^\beta_\mathrm{[KS]}}{\partial x^\nu_\mathrm{[BL]}}T^\mathrm{KS}_{\alpha\beta} ~,
\end{equation}
where
\begin{equation}
\frac{\partial x^\mu_\mathrm{[KS]}}{\partial x^\nu_\mathrm{[BL]}}=
\left[\begin{array}{cccc}
1 & \frac{2r}{\Delta} & 0 & 0 \\
0 & 1 & 0 & 0 \\
0 & 0 & 1 & 0 \\
0 & \frac{a}{\Delta} & 0 & 1
\end{array}\right] ~,
\end{equation}
with $\Delta=r^2-2r+a^2$.  For reference, we reproduce the basis vectors from \citet{Beckwith08}
\begin{eqnarray}
e^\mu_{\hat{r}} & = & -\frac{1}{k_rk_\phi}\left[\sqrt{g_{rr}}u^ru^t, \frac{k^2_\phi}{\sqrt{g_{rr}}}, 0, \sqrt{g_{rr}}u^ru^\phi\right] ~, \\
e^\mu_{\hat{\phi}} & = & -\frac{1}{k_\phi\sqrt{\vert-g^2_{t\phi}+g_{\phi\phi}g_{tt}\vert}}[g_{\phi\phi}u^\phi+g_{t\phi}u^t, 0, 0, -(g_{t\phi} u^\phi+g_{tt}u^t)] ~, \\
e^\mu_{\hat{\theta}} & = & -\frac{1}{k_rk_\theta}\left[\sqrt{g_{\theta\theta}}u^\theta u^t, \sqrt{g_{\theta\theta}}u^\theta u^r, \frac{k^2_r}{\sqrt{g_{\theta\theta}}}, \sqrt{g_{\theta\theta}}u^\theta u^\phi\right] ~,
\end{eqnarray}
where
\begin{eqnarray}
k_r & = & \sqrt{\vert g_{\phi \phi}(u^\phi)^2+g_{rr}(u^r)^2+u^t(2g_{t\phi}u^\phi+g^{tt}u^t)\vert} ~, \\
k_\theta & = & \sqrt{\vert g_{\phi \phi}(u^\phi)^2+g_{rr}(u^r)^2+g_{\theta\theta}(u^\theta)^2+u^t(2g_{t\phi}u^\phi+g^{tt}u^t)\vert} ~, \\
k_\phi & = & \sqrt{\vert g_{\phi \phi}(u^\phi)^2+u^t(2g_{t\phi}u^\phi+g^{tt}u^t)\vert} ~.
\end{eqnarray}
To calculate the rate-of-strain tensor components, we follow the same procedure, but replace the MHD stress-energy tensor (\ref{eq:tmunu}) with the covariant rate-of-strain tensor:
\begin{equation}
\sigma_{\alpha \beta}= \frac{1}{2}(u_{\alpha ; \mu} h^{\mu}_{\beta}
+ u_{\beta ; \mu} h^{\mu}_{\alpha}) - \frac{1}{3} u^{\mu}_{~;\mu} h_{\alpha \beta} ~, 
\label{eq:strain}
\end{equation}
where $h_{\alpha \beta} = g_{\alpha \beta} + u_\alpha u_\beta$ is the projection tensor.

%\bibliographystyle{apj}
%\bibliography{refs}

\end{document}